\documentclass[aps,prb,article,twocolumn,preprintnumbers,amsmath,amssymb,superscriptaddress]{revtex4-2}
\usepackage{float}
\usepackage{commath}
\usepackage{graphicx}
\usepackage{verbatim}
\usepackage{color}
\usepackage{physics}
\usepackage[dvipsnames,svgnames,x11names,hyperref]{xcolor}
\usepackage[normalem]{ulem}
\usepackage{algorithm}
\usepackage{algpseudocode}
\usepackage{times}
\usepackage[compat=1.1.0]{tikz-feynman}\tikzfeynmanset{warn luatex=false}
\usepackage[toc,page]{appendix}
\usepackage{siunitx}
\usepackage{dcolumn}   
\usepackage{bm}        
\usepackage{amssymb}   
\usepackage{amsmath}
\usepackage{epigraph}
\usepackage{braket}
\usepackage{gensymb}
\usepackage{ tipa }
\usepackage{bbold}
\usepackage{esint}
\usepackage{mathdots}
\usepackage{float}
\usepackage{comment}
\usepackage[colorlinks=true, linkcolor=blue, urlcolor=blue, citecolor=blue]{hyperref}
\DeclareSIUnit\gauss{G}

\definecolor{linkcolor}{RGB}{6,69,173} 
\definecolor{diffcolor}{RGB}{175,31,36} 

\newcommand{\Ham}{\mathcal{H}}

\newcommand{\eps} {{\bm{\varepsilon}}}

\newcommand{\epsin} {{\eps_{\rm i}}}
\newcommand{\epsout} {{\eps_{\rm f}}}
\newcommand{\win} {{\omega_{\rm in}}}

\newcommand{\dw} {{\omega}}
\newcommand{\Adagger}{A^{\dagger}}

\allowdisplaybreaks

\begin{abstract}
Resonant spectroscopies, which involve intermediate states with finite lifetimes, provide essential insights into collective excitations in quantum materials that are otherwise inaccessible. However, theoretical understanding in this area is often limited by the numerical challenges of solving Kramers-Heisenberg-type response functions for large-scale systems. To address this, we introduce a multi-shifted biconjugate gradient algorithm that exploits the shared structure of Krylov subspaces across spectra with varying incident energies, effectively reducing the computational complexity to that of linear spectroscopies. Both mathematical proofs and numerical benchmarks confirm that this algorithm substantially accelerates spectral simulations, achieving constant complexity independent of the number of incident energies, while ensuring accuracy and stability. This development provides a scalable, versatile framework for simulating advanced spectroscopies in quantum materials.
\end{abstract}

\begin{document} 
\title{Accelerating Resonant Spectroscopy Simulations Using Multi-Shifted Bi-Conjugate Gradient}
\author{Prakash Sharma}
\affiliation{Department of Chemistry, Emory University, Atlanta, GA 30322, United States}
\author{Luogen Xu}
\affiliation{Department of Chemistry, Emory University, Atlanta, GA 30322, United States}
\author{Fei Xue}
\affiliation{School of Mathematical and Statistical Sciences, Clemson University, Clemson, SC 29634, United States}
\author{Yao Wang}
\email{yao.wang@emory.edu}
\affiliation{Department of Chemistry, Emory University, Atlanta, GA 30322, United States}

\maketitle

\section{Introduction}
Understanding how collective excitations---such as spin, charge, orbital, and lattice modes---interact and evolve across energy scales is central to revealing the emergent properties and functionalities of quantum materials\,\cite{basov2017towards, keimer2017physics}. These excitations are typically probed using momentum- and energy-resolved spectroscopic techniques, such as optical absorption, Raman, and neutron scattering. As research has deepened, increasing attention has been drawn to collective modes like selection-rule-forbidden orbital excitations, entanglement, or hidden orders. These challenges have spurred the rapid development of nonlinear resonant spectroscopies over the past two decades\,\cite{de2024resonant,mitrano2024exploring,mitrano2020probing,liu2025entanglement}. A particularly powerful technique is the resonant inelastic x-ray scattering (RIXS)\,\cite{kotani2001resonant,ament2011resonant}, which has enabled groundbreaking discoveries in cuprates\,\cite{ghiringhelli2012long, dean2013persistence, hepting2018three}, nickelates\,\cite{lu2021magnetic, gao2024magnetic,chen2024electronic}, and other transition-metal compounds\,\cite{kim2012magnetic,pelliciari2021evolution,zhang2019trace,mahmood2022distinguishing}. Other resonant spectra, such as two-photon absorption, resonant Raman scattering and pair photoemission, also provide crucial information of unconventional symmetries and fluctuations in quantum materials\,\cite{peticolas1963double,devereaux2007inelastic, ko2010raman,de2020effective, devereaux2023angle, hsu2025detection}.

\begin{figure}[!t]
\begin{center}
    \includegraphics[width=\linewidth]{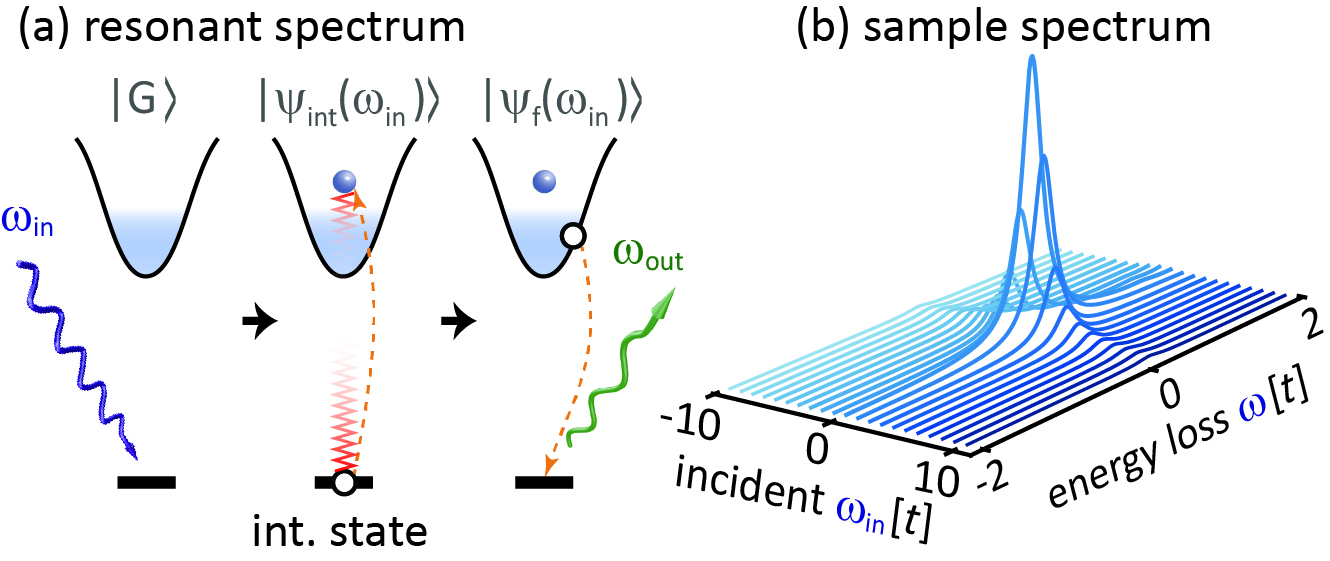}\vspace{-2mm}
    \caption{\label{fig:RIXSCartoon}
    (a) Schematic illustrating a typical resonant scattering process, where an intermediate state with finite lifetime and many-body effects is generated. (b) Example of a resonant spectrum, the spin-flip RIXS response for magnon excitations, showing the dependence of scattering intensity on both incident photon energy and energy loss.}
\end{center}
\end{figure}

Efficiently using these spectral techniques requires quantitatively linking the response functions to specific underlying excitations. While linear spectroscopies can often be unified under two-point dynamical correlations governed by the Kubo formalism, resonant spectroscopies require higher-order perturbative treatments described by the Kramers-Heisenberg formalism\,\cite{ament2011resonant}. This response function explicitly incorporates the dynamics of the intermediate state\,\cite{wang2018theoretical}. Typically,  an incident photon excites the system into an intermediate state $\ket{\Psi_{\rm int}}$ with a finite lifetime (denoted as $1/\Gamma$), after which the system radiatively relaxes by emitting a photon [see Fig.~\ref{fig:RIXSCartoon}(a)]. The cross-section for this two-step process is calculated as
\begin{equation}\label{eq:kramers1}
    I\left(\win,\dw\right) \propto\textrm{Im}\bra{\Psi_{\rm int}}\mathcal{\hat{O}}^\dagger_{1}\frac1{\mathcal{H}-E_G-\dw-i0_+} \mathcal{\hat{O}}_{1}\ket{\Psi_{\rm int}},
\end{equation}
with the intermediate state constructed by
\begin{equation}\label{eq:intermediateState}
    \ket{\Psi_{\rm int}(\win)} = \frac{1}{\Ham' -E_G - \win - i\Gamma} \mathcal{\hat{O}}_{2}\ket{G}\,.    
\end{equation}
Here, $E_G$ and $\ket{G}$ denote the ground-state energy and wavefunction; $\win$ and $\win-\dw$ are the photon energies involved in the two steps of the response, and $\Ham^\prime$ is the Hamiltonian governing the intermediate state, which may differ significantly from the initial-state Hamiltonian $\Ham$. The operators $\mathcal{\hat{O}}_{1/2}$ encode the relevant excitations: for RIXS, they are dipole transition operators; for Raman scattering, they are symmetry-resolved current operators; and for other resonant spectroscopies, they are defined accordingly.

At the ultrashort lifetime limit ($\Gamma\rightarrow\infty$), the intermediate state becomes virtual, and Eq.~\eqref{eq:kramers1} simplifies to the Kubo formalism\,\cite{ ament2009theoretical,van2007theory,braicovich2009dispersion,forte2008magnetic}, allowing the full energy spectrum to be computed efficiently using the continued fraction expansion\,\cite{dagotto1994correlated}. However, this approximation fails to reflect the realistic energy scales and spectral distributions and can significantly underestimate the contributions of critical excitations\,\cite{jia2016using}. As a result, obtaining realistic predictions requires explicit solution of the full Kramers-Heisenberg response, i.e., Eqs.~\eqref{eq:kramers1} and \eqref{eq:intermediateState}. This simulation challenge arises because calculating the cross-section requires evaluating the intermediate state $\ket{\Psi_{\rm int}(\win)}$ for each incident energy $\win$ [see Fig.~\ref{fig:RIXSCartoon}(b)]. Each such evaluation relies on solving a linear system defined by Eq.~\eqref{eq:intermediateState}, and for quantum many-body systems, where the Hilbert space dimension grows exponentially with system size, this step constitutes the dominant computational bottleneck. Thus, the overall computational cost of simulating resonant spectroscopies is vastly higher than that of linear spectroscopies, and the cost scales linearly with the number of incident energies sampled. This severe computational complexity has significantly limited the scope and accuracy of resonant spectroscopy simulations, posing a major challenge for theoretical and computational studies in the field.

To address the computational bottlenecks inherent in simulating Kramers-Heisenberg-type resonant spectroscopies, we introduce an algorithm based on the multi-shifted biconjugate gradient (MSBiCG). This method leverages the invariance of Krylov subspaces under constant shifts and the colinearity of residuals, enabling the reuse of Krylov vectors across different incident energies. By eliminating redundant matrix-vector multiplications (MVMs) typically required to construct separate Krylov subspaces, the computational cost becomes independent of the number of $\win$s, reducing the computational cost to a level comparable with that of linear spectral simulations. This advance enables efficient, scalable numerical simulations of large-scale resonant spectroscopies.

\section{Multi-shifted Biconjugate Gradient}
\begin{figure}[!t]
\begin{center}
    \includegraphics[width=\linewidth]{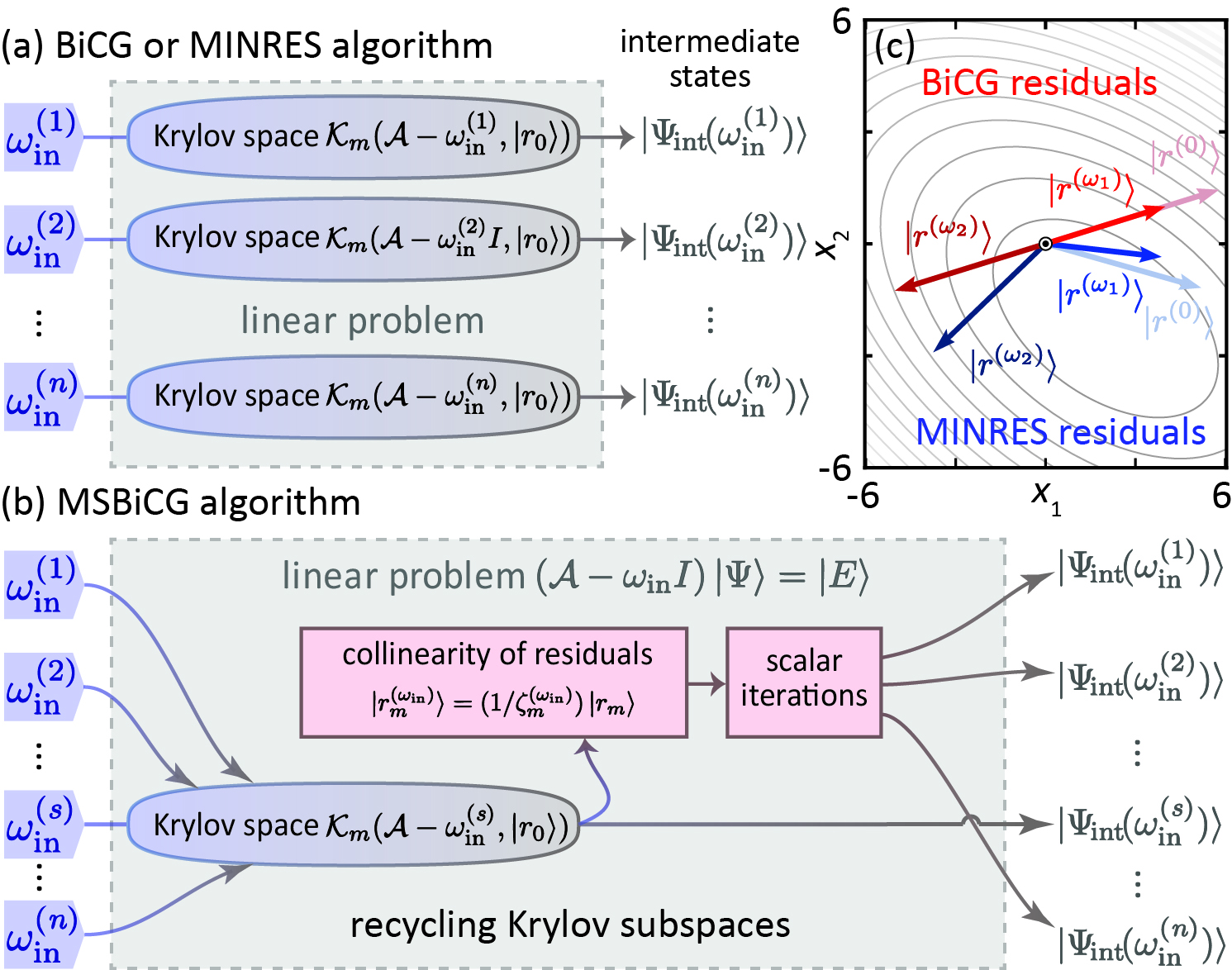}
    \caption{\label{fig:twoLinearSolvers}
    (a) Flowchart illustrating the BiCG or MINRES algorithms for solving the linear problem in Eq.~\eqref{eq:intermediateState} across multiple incident energies, where each incident energy is treated as an independent problem, requiring separate Krylov space construction. (b) Flowchart illustrating the MSBiCG algorithm, which utilizes the collinear residuals and recycles Krylov spaces generated from the seed system. (c) Example using a $2 \times 2$ toy matrix system, illustrating the collinearity of BiCG residuals (red arrows) in a two-dimensional vector space for different shifts. In contrast, residuals from MINRES (blue arrows) do not satisfy this collinearity condition. The contour represents the quadratic form, $\bra{\Psi}\mathcal{A}\ket{\Psi}/2-\langle E |\Psi\rangle$ for this toy system. } 
\end{center}
\end{figure}

To set the stage for our approach, we first review how the resonant spectroscopies described by Eq.~\eqref{eq:kramers1} are traditionally addressed in numerical methods. Because direct matrix inversion is numerically unstable, the propagator in Eq.~\eqref{eq:intermediateState} is reformulated as the linear problem $(\mathcal{A}-\win I)\ket{\Psi_{\rm int}}=\ket{E}$, with $\mathcal{A}=\Ham'-E_G-i\Gamma \in \mathbb C^{D\times D}$ and $\ket{E}=\mathcal{\hat{O}}_{2}\ket{G} \in \mathbb C^{D}$. Both the (sparse) matrix $\mathcal{A}$ and the excited-state wavefunction $\ket{E}$ reside in a Hilbert space of dimension $D$. Solving such large sparse problems efficiently relies on Krylov-subspace methods, which construct the $m$-dimensional Krylov space:
\begin{equation}
    \mathcal K_m (\mathcal{A}, \ket{r_0}) = \mathrm{span} \{\ket{r_0}, \mathcal{A}\ket{r_0}, \cdots, \mathcal{A}^{m-1}\ket{r_0} \}\,.
\end{equation}
Here, $\ket{r_0}=\ket{E}-\mathcal{A}\ket{\Psi_0}$ is the initial residual with initial guess $\ket{\Psi_0}$ (see Appendix \ref{App:biorthogonality}). Among Krylov approaches, minimal residual (MINRES)\,\cite{wang2019edrixs} and biconjugate gradient stabilized (BiCGStab) algorithms\,\cite{chen2010unraveling,jia2012uncovering,jia2014persistent,jia2016using} are widely employed in exact diagonalization (ED) calculations of resonant spectra due to their numerical stability\,\cite{biorthogonal}; conjugate gradient and correction-vector method are implemented with the density matrix renormalization group (DMRG) for simulations in long 1D chains\,\cite{kuhner1999dynamical, nocera2016spectral,kumar2018multi,nocera2018computing}. However, these advanced techniques face the aforementioned bottlenecks when covering various excitations across a wide range of incident energies. This challenge arises because solving Eq.~\eqref{eq:intermediateState} iteratively demands constructing a separate Krylov subspace for each $\omega_\mathrm{in}$ [see Fig.~\ref{fig:twoLinearSolvers}(a)], resulting in significantly more MVMs compared to linear ones solvable by the continued fraction expansion\,\cite{dagotto1994correlated}.

Rather than treating each individual incident energy as an independent problem, we recognize that the systems differ only by a constant shift in the matrix. As a result, the Krylov subspace of a shifted system, $\mathcal{K}_m(\mathcal{A}-\win I,\ket{E})$, is the same Krylov subspace of a pre-selected seed system, $\mathcal{K}_m(\mathcal{A},\ket{E})$, if $\ket{\Psi_0}=\ket{0}$ (null vector) is used for all systems [see the Appendix~\ref{App:collinearity} for a detailed proof]. This observation motivates a strategy to recycle Krylov vectors across all shifted problems (corresponding to spectra at different $\win$s) simultaneously, eliminating the need for additional MVMs. 
  
This strategy achieves particular efficiency when combined with iterative methods satisfying the Petrov-Galerkin condition, such as CG and BiCG\,\cite{van1990petrov}. In these methods, not only are the Krylov subspaces identical across all shifted systems, but the individual residual vectors at each iteration are guaranteed to be collinear across shifts (see Appendix~\ref{App:collinearity} for a brief proof and Refs.~\onlinecite{frommer2003BiCGStab, meng2015recycling} for a full mathematical proof):
\begin{equation}\label{eq:shiftCoeff}
    \ket{r_m^{(\win)}} = (1/\zeta_m^{(\win)})\ket{r_m},
\end{equation}
where $\ket{r_m^{(\win)}}$ is the $m$-th residual for the shifted system with incident energy $\win$ and $\ket{r_m}$ is the residual of the seed system. This collinearity ensures that, as long as the same initial guess $\ket{\Psi_0}=\ket{0}$ is used for all shifted problems, the algorithm only needs to track the scalar coefficient $\zeta_m^{(\win)}$ at each iteration for each shifted system, avoiding additional vector operations. By contrast, methods like MINRES do not satisfy the collinearity condition and therefore require extra vector calculations for each shift, as illustrated in Fig.~\ref{fig:twoLinearSolvers}(c).

Following the standard three-term recurrence for the BiCG residual and its polynomial form, the recurrence relation for the scaling factors reads as (see details in Appendix~\ref{App:Iterative_eq}):
\begin{equation}\label{eq:collinear_recursion}
    \zeta_{m+1}^{(\win)} = \left(1+\alpha_m \win \right)\zeta_m^{(\win)} + \frac{\alpha_m\beta_{m}}{\alpha_{m-1}} \left( \zeta_m^{(\win)} - \zeta_{m-1}^{(\win)} \right)
\end{equation}
where $\alpha_m={\bra{r_m}r_m\rangle}/{\bra{r_m}\mathcal{A}\ket{p_m}}$ ($\ket{p_m}$ is search direction) and $\beta_m={\bra{r_{m+1}}r_{m+1}\rangle}/{\bra{r_{m}}r_{m}\rangle}$ are the Lanczos coefficients of the seed system. (Here, we have avoided introducing dual vectors, as explained in Appendix~\ref{App:biorthogonality}.) Once the BiCG residual $\ket{r_m}$ (and its corresponding $\alpha_m$ and $\beta_m$) is computed for the seed system, the residuals for all shifted systems can be updated directly using Eqs.~\eqref{eq:shiftCoeff} and \eqref{eq:collinear_recursion}.

With the updated $\zeta_{m}^{(\win)}$ and the residual vectors $\ket{r_m^{(\win)}}$, we can calculate the $\win$-specific BiCG coefficients $\alpha_m^{(\win)}$ and $\beta_m^{(\win)}$ without performing any additional MVMs. Thanks to the collinearity property, these coefficients are straightforwardly scaled from the seed-system ones:
\begin{equation}
\alpha_m^{(\win)} = \left(\frac{\zeta_m^{(\win)}}{\zeta_{m+1}^{(\win)}}\right) \alpha_m \text{ }\text{ and }\text{ } \beta_m^{(\win)} = \left(\frac{\zeta_{m-1}^{(\win)}}{\zeta_m^{(\win)}}\right)^2 \beta_m
\label{eq:coeffs}
\end{equation}
Finally, the search directions and approximate solutions for the shifted systems are updated analytically (see Appendix~\ref{App:Iterative_eq} for the derivations and the full algorithm in Appendix~\ref{App:MSBICG_algorithm}):
\begin{align}
\ket{p_m^{(\win)}} &= \left(1/ \zeta_m^{(\win)} \right)\ket{r_m} + \beta_m^{(\win)} \ket{p_{m-1}^{(\win)}}
\label{eq:search_dir} \\
 \ket{\Psi_{m+1}^{(\win)}} &= \ket{\Psi_{m}^{(\win)}} + \alpha_m^{(\win)} \ket{p_m^{(\win)}}.
\label{eq:solution_ket}
\end{align}
After iterating, we monitor the convergence of both the seed and all shifted systems, terminating the algorithm once all systems have converged.

Although the iterative steps in Eqs.~\eqref{eq:shiftCoeff} through \eqref{eq:solution_ket} involve multiple vector and scalar operations, their computational cost is negligible for large systems since no MVM is conducted. Consequently, the overall computational complexity of the MSBiCG algorithm is effectively determined by the MVM count of the seed system, yielding $(m - 1) $ for any resonant spectral calculations, regardless of the number of incident energies involved. For systems with complex Hamiltonian entries, such as a dissipative system or a model breaking time-reversal symmetry, the need to explicitly construct the dual subspace doubles the MVM count\,\cite{biorthogonal}. In contrast, a conventional BiCGStab algorithm incurs $2n(m - 1)$ MVMs when treating $n$ separate $\win$s, reflecting the substantial performance acceleration offered by MSBiCG. 

\begin{figure*}[t!]
    \centering
    \includegraphics[width=\textwidth]{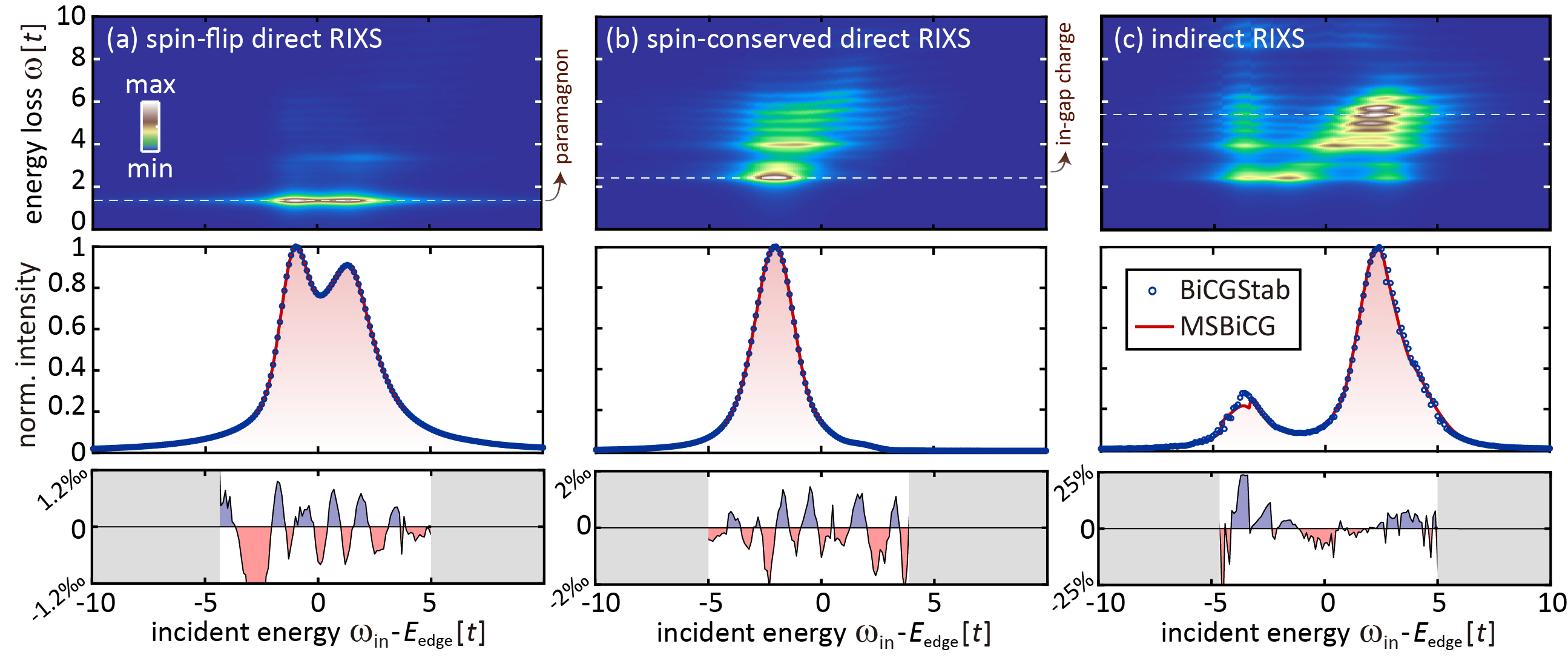}\vspace{-3mm}
    \caption{\label{fig:singleband_intensity}
    (a) Top: RIXS spectra simulated using the MSBiCG algorithm for the spin-flip channel at momentum transfer $\mathbf{q}=(\pi,\pi)$, based on a 12.5\% doped Hubbard model and with inverse core-hole lifetime $\Gamma = t$. Middle: Comparison between simulated spectra obtained using BiCGStab (blue circles) and MSBiCG (red line) along the $\omega = 1.35t$ cut, indicated by the dashed line in the top panel. Bottom: Relative spectral error (near the resonance peak) along this cut as a function of $\win$s. (b) Same as (a) but for the spin-conserved channel and simulated in a 25\% doped Hubbard model. (c) Same as (b) but for indirect RIXS simulations.} 
\end{figure*}

\section{Performance Benchmark in Hubbard-model Systems}

To benchmark the performance of the MSBiCG algorithm on a physically relevant system, we simulate RIXS spectra for a single-band Hubbard model, a representative model for strongly correlated electrons, using both the BiCGStab and MSBiCG algorithms. The Hamiltonian is given as
\begin{equation}\label{eq:Hubbard}
    \mathcal H = -\sum_{i,j, \sigma}t_{ij}(c_{j\sigma}^\dagger c_{i\sigma}+h.c) + U\sum_i n_{i\uparrow}n_{i\downarrow},
\end{equation}
where $c_{i\sigma}^\dagger$ ($c_{i\sigma}$) creates (annihilates) an electron with spin $\sigma$ at site $i$, and $n_{i\sigma} = c_{i\sigma}^\dagger c_{i\sigma}$ denotes the local density. We truncate the hopping integrals $t_{ij}$ to the nearest neighbor $t=0.4\,$eV and next-nearest neighbor  $t^\prime=-0.3t$, with the on-site Coulomb interaction fixed at $U = 8t$. The ground state $\ket{G}$ is calculated using parallel Arnoldi method with Paradeisos acceleration\,\cite{lehoucq1998arpack,  jia2017paradeisos}.  

The intermediate state $\ket{\Psi_{\rm int}}$ is generated by substituting the generic operators $\mathcal{\hat{O}}_{1}$ and $\mathcal{\hat{O}}_{2}$ in Eqs.~\eqref{eq:kramers1} and \eqref{eq:intermediateState} with the specific dipole transition operators $\mathcal{D}_{j\epsout}^\dagger$ and $\mathcal{D}_{j\epsin}$ relevant to the chosen RIXS process\,\cite{ament2011resonant}. For example, at a direct edge such as the Cu $L$-edge, the dipole operator is expressed as $\mathcal{D}_{i\epsin} = \sum_{\mu\sigma}M^{(\epsin)}_{\mu\sigma}c^\dagger_{i\sigma}h_{i\mu\sigma}$, where $h_{i\mu\sigma}$ annihilates a core-level ($2p_{\mu}$ orbital) electron at site $i$ and the transition matrix elements are given by $M_{\mu\sigma}^{(\epsin)}=\bra{3d}\hat{\mathbf{\epsin}}\cdot \hat{\mathbf{r}}\ket{2p_\mu}$\,\cite{jia2014persistent, jia2016using}. The introduction of a core hole modifies the system’s Hamiltonian from Eq.~\eqref{eq:Hubbard}, adding interaction terms between the core hole and the valence electrons:
\begin{equation}\label{eq:intermediate_H}
    \mathcal{H}'\mkern-2mu = \mathcal{H}+E_{\rm edge}\mkern-2mu\sum_{i\mu\sigma} h_{i\mu\sigma} h_{i\mu\sigma}^\dagger\mkern-2mu-U_c\mkern-4mu\sum_{i\mu\sigma \sigma'} n_{i\sigma} h_{i\mu\sigma'} h_{i\mu\sigma'}^\dagger\,.
\end{equation}
In this paper, we set the core-hole interaction $U_c=4t$\,\cite{jia2016using}.

Figure \ref{fig:singleband_intensity}(a) presents MSBiCG-based RIXS simulations for a $\pi$-$\sigma$ polarization configuration at momentum transfer $\mathbf{q}=(\pi,\pi)$ applied to a $12.5\%$ hole-doped 16B Betts cluster\,\cite{betts1999improved}, designed to highlight the spin-flip channel. The intermediate state’s dimension in this system reaches approximately $10^9$. Without loss of generality, we select the atomic resonance energy $\win=E_{\rm edge}$ as the seed system, allowing Eqs.~\eqref{eq:collinear_recursion}-\eqref{eq:solution_ket} to generate solutions across all other incident energies. As depicted in Fig.~\ref{fig:RIXSCartoon}(a), the simulated RIXS spectrum prominently reveals a paramagnon excitation spanning $1.3t$ to $3t$. To visualize the accuracy of MSBiCG, we extract an incident-energy cut at $\omega = 1.35t$ and scan over 200 different $\win$s (see the middle panel). The RIXS intensity computed by MSBiCG shows excellent agreement with that obtained from independent BiCGStab runs for each $\win$. Notably, near the spectral peak, where precision matters most, the relative discrepancy between the two methods is constrained to below $0.1$\% (see the corresponding bottom panel).

Similarly, we simulate the RIXS spectrum using a different polarization configuration while maintaining the same model parameters, now at $25\%$ hole doping. To provide a complementary comparison to Fig.~\ref{fig:singleband_intensity}(a), we focus exclusively on the spin-conserved channel to compare, where the intermediate-state Hilbert space dimension is $10^8$. The spectrum exhibits a single resonance at $\win\sim -2t$, whose excitation distribution physically captures the in-gap charge mode induced by doping\,\cite{wang2020emergence}. When comparing the spectral cut at $\omega=2.4t$, we again observe an excellent agreement between the two algorithms, with the maximal relative error under $0.2\%$.

While the previous examples focused on direct RIXS spectra, an equally important class of processes is indirect RIXS\,\cite{ament2011resonant}, which is triggered by x-ray absorption into higher-energy orbitals rather than the valence shell. Simulating indirect RIXS requires substituting the valence operators in Eq.~\eqref{eq:intermediate_H} and the dipole excitation into electronic terms with electronic operators acting on higher empty orbitals (e.g.~the 4$p$ orbital for a Cu $K$-edge RIXS). Detailed descriptions of this model are provided in Refs.\,\onlinecite{jia2012uncovering, van2005correlation, van2007theory} and are not repeated here. For completeness, we benchmark the MSBiCG algorithm for indirect RIXS using the same model parameters as Fig.~\ref{fig:singleband_intensity}(b). As shown in Fig.~\ref{fig:singleband_intensity}(c), the simulated spectrum features two main peaks, corresponding to the poorly-screened and hole-doped absorptions\,\cite{jia2012uncovering}, respectively (a third resonance near $\win = E_{\rm edge}$, representing the well-screened absorption, is present but weak and unresolved under current parameters). The simulation results from the two algorithms align well for the main peak at $\win\sim2.4t$, yielding a relative error of just $1.1\%$. However, the MSBiCG algorithm shows larger discrepancies at the other resonant peak near $\win\sim-3.6t$, with the relative error increasing to $23\%$, as the residuals fail to reach the convergence threshold of $10^{-6}$, within the fixed iteration limit of 1000. This behavior reflects a limitation of MSBiCG when multiple strong poles are distributed across $\win$. In such cases, the Krylov subspaces of shifted systems are generated through the recursions in Eqs.\eqref{eq:collinear_recursion} and \eqref{eq:coeffs}, and round-off errors accumulate as iterations proceed. These errors strongly affect accuracy at distant resonances, particularly when they require many iterations to resolve. For spectra covering a broad incident-energy range with multiple competing resonances, this issue can be systematically addressed using an adaptive reseeding strategy [see Appendix~\ref{App:adaptive_reseeding} for details].

\begin{figure}[t!]
    \centering  \includegraphics[width=\linewidth]{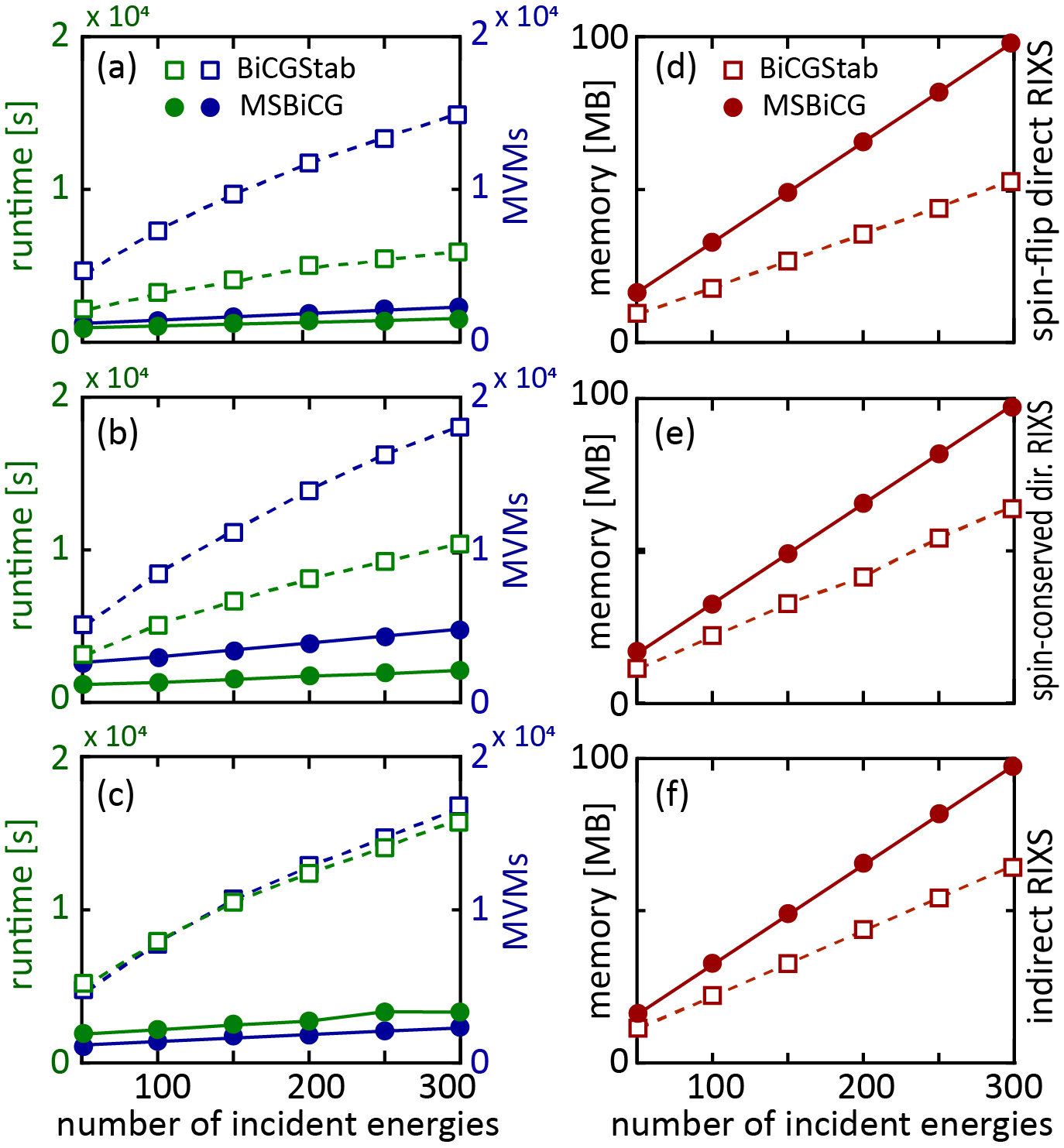}\vspace*{-2mm}
    \caption{(a-c) Runtime (green) and total MVM counts (blue) comparison between BiCGStab (open squares) and MSBiCG (filled circles) as a function of the total number of incident energies, for the RIXS simulations shown in Figs.~\ref{fig:singleband_intensity}(a-c), respectively. All simulations use a residual threshold of $10^{-6}$. (d-f) Memory consumption for the corresponding RIXS simulations, comparing BiCGStab (open squares) and MSBiCG (filled circles). }
    \label{fig:sim_cost}
\end{figure}

After assessing accuracy, we turn to benchmarking the efficiency of the MSBiCG algorithm relative to BiCGStab. Figs.~\ref{fig:sim_cost}(a)-(c) compare runtime and total MVMs for both algorithms across various RIXS simulation cases. As discussed above, the theoretical time complexity of MSBiCG scales as $(m-1)$, in contrast to the $2n(m-1)$ of BiCGStab. In practice, however, MSBiCG often requires more iterations, compared to a single BiCGStab problem, to reach the same convergence threshold across all $\win$s, as performance is governed by the most ill-conditioned shifted system (typically near resonance). As a result, the observed acceleration is often well below the ideal $2n$ factor, especially for relatively small $n$s\,\cite{BiCGStab}. Nevertheless, as the density of incident energies increases, the acceleration becomes increasingly pronounced: for instance, on a dense meshgrid with 300 $\win$s, MSBiCG outperforms BiCGStab by roughly an order of magnitude. More importantly, the scaling matches theoretical expectations: the runtime and MVM count of BiCGStab increase roughly linearly with the number of $\win$s, whereas those of MSBiCG remain largely constant. The small upward deviation in MSBiCG’s cost arises from the additional Krylov iterations required to maintain the same residual convergence as the number of incident energies increases.

The MSBiCG algorithm saves iteration time by recycling the Krylov subspace but does so at the expense of increased memory usage, as it must store all shifted residual vectors from Eq.~\eqref{eq:shiftCoeff}. To assess this memory overhead, we analyze the storage requirements for the MSBiCG and BiCGStab algorithms across the simulations [see Figs.~\ref{fig:sim_cost}(d)-(f)]. As expected, the memory cost for MSBiCG increases linearly with the number of incident energies, reflecting the added storage for shifted vectors, and grows more steeply than BiCGStab. It is important to highlight that our BiCGStab-based spectral simulation employs a specific time-efficient approach: we first solve all intermediate states $|\Psi_{\rm int}(\win)\rangle$s and only then compute the final-state spectrum using Eq.~\eqref{eq:kramers1} via continued fraction expansion. This strategy avoids simultaneously storing both the $\Ham$ and $\Ham'$ Hamiltonian matrices in memory and eliminates the need to repeatedly reconstruct the final-state Hamiltonian. For extremely large numbers of $\win$s, an alternative strategy of retaining both Hamiltonians in memory could be considered, which would cap the overall memory cost by allowing computed $|\Psi_{\rm int}(\win)\rangle$ to be released. However, we do not address this extreme case within the scope of this paper, assuming instead that minimizing Hamiltonian memory costs remains the priority. 
 
\section{Conclusion}

We have developed the musti-shifted biconjugate gradient (MSBiCG) algorithm for solving general resonant spectroscopies within the Kramers-Heisenberg formulism. By exploiting the Krylov subspace structure across varying incident energies, this method achieves nearly constant computational complexity with respect to the number of incident photon frequencies, reducing the simulation cost to a level comparable to that of linear response calculations. MSBiCG thus provides a highly efficient alternative to standard linear solvers such as BiCGStab and MINRES, substantially lowering computational overhead for complex spectroscopies. While our demonstrations focus on RIXS, the algorithm is broadly applicable to a wide range of resonant spectroscopies, including two-photon absorption, resonant Raman scattering, and pair photoemission. Furthermore, the MSBiCG framework can be integrated with tensor network methods, such as DMRG, offering the potential to expand the simulation capabilities for spectroscopies in large-scale quantum systems. 

\section*{Acknowledgments}

We acknowledge helpful discussions with J. Hales. This work is primarily supported by the U.S. Department of Energy, Office of Science, Basic Energy Sciences, under Early Career Award No.~DE-SC0024524. F.X. acknowledges the support by the U.S. National Science Foundation, under award DMS-2111496. The simulation used resources of the National Energy Research Scientific Computing Center, a U.S. Department of Energy Office of Science User Facility located at Lawrence Berkeley National Laboratory, operated under Contract No.~DE-AC02-05CH11231.

\section*{Data Availability}

The data that support the findings of this article are available in a public repository at \onlinecite{data}.

\appendix

\section{Simplification of the Biorthogonality Condition}\label{App:biorthogonality}
The intermediate-state problem involved in a resonant spectrum, described by Eq.~\eqref{eq:intermediateState}, maps to the linear equation 
\begin{align}
     \left(\mathcal{A}-\win I \right)\ket{\Psi_{\rm int}(\win)} 
    &= \ket{E}, \label{eq:shifted_linear_eq}
\end{align}
where $\mathcal{A} = \left(\Ham' -E_G - i\Gamma \right)$ and $\ket{\Psi_{\rm int}(\win)}$ is the intermediate state induced by the $\win$-energy of incident photon. The linear solvers discussed in this paper focus on efficiently solving $\ket{\Psi_{\rm int}(\win)}$s from a sequence of Eq.~\eqref{eq:shifted_linear_eq} with various $\win$s using $m$-dimensional Krylov subspace:
\begin{equation}
    \mathcal K_m (\mathcal{A}, \ket{r_0}) = \mathrm{span} \{\ket{r_0}, \mathcal{A}\ket{r_0}, \cdots, \mathcal{A}^{m-1}\ket{r_0} \}\,.
\end{equation}
In the context of general linear systems involving non-Hermitian operators—such as our matrix $\mathcal{A}$—the inherent asymmetry of $\mathcal{A}$ leads to a breakdown of orthogonality among Krylov vectors. To address this issue, the biconjugate gradient (BiCG) method requires the dual subspace of $\mathcal{A}$, i.e., 
\begin{equation}
\mathcal L_m = \mathrm{span}\{ \ket{\tilde{r}_0}, \mathcal{\Adagger} \ket{\tilde{r}_0} , \cdots,(\mathcal{\Adagger})^{m-1} \ket{\tilde{r}_0} \},
\end{equation}
in order to maintain the biorthogonality conditions between the primal and dual vector spaces, namely
\begin{equation}
    \bra{\tilde{r}_j} r_i \rangle = \delta_{ij}
    \quad \text{ and  }\quad
    \bra{\tilde{p}_j} \mathcal{\Adagger}\ket{p_i}  = \delta_{ij}
    \label{eq:bicg_orth},
\end{equation}
where $\ket{r_i}$ 
($\ket{\tilde{r}_i}$) is the $i^{\rm th}$ residual and $\ket{p_i}$ ($\ket{\tilde{p}_i}$) is the search direction constructed using the matrix $\mathcal{A}$ ($\mathcal{\Adagger}$), and the initial dual space residual is $\ket{\tilde{r}_0}=\ket{b}-\mathcal{\Adagger}\ket{\Psi_0}$. Eqs.~\eqref{eq:bicg_orth} are an essential condition for ensuring the stability and convergence in BiCG method.  
However, constructing the dual subspace $\mathcal{L}_m$ is just as computationally demanding as generating the primary Krylov subspace $\mathcal{K}_m$. Fortunately, for most systems of physical interest, the Hamiltonian $\Ham'$ is Hermitian and can be written as a real-symmetric matrix in a certain basis, making $\mathcal{A}=\Ham' -E_G  - i\Gamma$ complex only along its diagonal. In such cases, the explicit construction of $\mathcal{L}_m$ is no longer required; rather, we can use the Hermitian conjugate of the seed residuals in its place, i.e., $\ket{\tilde{r}_i}=(\ket{r_i})^*$, thereby relaxing the biorthogonality condition. 

To demonstrate this, let us consider the seed system $\mathcal{A}\ket{\Psi}=\ket{E}$ with the initial guess solution chosen as the null vector, $\ket{\Psi_0}=\ket{0}$. Although this initial choice is not strictly required here, later we will see this choice plays important role in maintaining collinearity among residuals, when shifted systems are also considered. With the chosen initial condition, the initial residuals, for $m=0$,  simplify to 
\begin{align}
    \ket{r_0} &= \ket{E}-\mathcal{A}\ket{\Psi_0} =\ket{E}
    \label{eq:seed_init_residual}
    \\
    \ket{\tilde{r}_0} &= \ket{E}-\mathcal{A}^\dagger\ket{\Psi_0} =\ket{E}.
    \label{eq:seed_init_dualresidual}
\end{align}
This automatically satisfies $\ket{\tilde{r}_0}=\ket{r_0}^*$ since $\ket{E}\in \mathbb{R}^D$.

For $m=1$, the approximate solution updates in the first Krylov iteration are given by $\ket{\Psi_1}=\ket{\Psi_0}+\alpha_0\ket{r_0}=\alpha_0\ket{E}$ and $\ket{\tilde{\Psi}_1}=\alpha_0^*\ket{E}$ for some scalar coefficient $\alpha_0\in \mathbb{C}$. The the corresponding residuals are
\begin{align}
     \ket{r_1} &= \ket{E}-\mathcal{A}\ket{\Psi_1}= \ket{E}-\alpha_0\mathcal{A}\ket{E} 
     \label{eq:r1_tilde}
    \\
    \ket{\tilde{r}_1} &=  \ket{E}-\mathcal{A}^\dagger \ket{\tilde{\Psi}_1} = \ket{E}-\alpha_0^*\mathcal{A}^\dagger\ket{E}\,.
    \label{eq:r1_tilde_dual}
\end{align}
Since the Hamiltonian $\mathcal{H}'\in \mathbb{R}^{D\times D}$ is real symmetric matrix, $\mathcal{A}^\dagger=\Ham' -E_G  + i\Gamma =\mathcal{A}^*$.  Therefore, from Eqs.~\eqref{eq:r1_tilde} and \eqref{eq:r1_tilde_dual}, we have 
\begin{equation}  
\ket{\tilde{r}_1} = \ket{E}-\alpha_0^*\mathcal{A}^*\ket{E}=\ket{r_1}^*
\end{equation} 
Now let us suppose $\ket{\tilde{r}_m}=\ket{r_m}^*$. 
Then, the $(m+1)^\text{th}$ residuals satisfy
\begin{align}\label{eq:rm_update}
  \ket{r_{m+1}} 
    &= \ket{E} - \mathcal{A} \ket{\Psi_{m+1}} = \ket{E} - \mathcal{A}(\ket{\Psi_m} + \alpha_m \ket{r_m}) \nonumber \\
    &= \ket{r_m} - \alpha_m \mathcal{A} \ket{r_m} \,,
\end{align}
and
\begin{align}\label{eq:rtilde_update}
  \ket{\tilde{r}_{m+1}} 
    &= \ket{E} - \mathcal{A}^\dagger \ket{\tilde{\Psi}_{m+1}} = \ket{E} - \mathcal{A}^\dagger(\ket{\tilde{\Psi}_m} + \alpha_m^* \ket{\tilde{r}_m}) \nonumber \\
    &= \ket{\tilde{r}_m} - \alpha_m^* \mathcal{A}^\dagger \ket{\tilde{r}_m} \,.
\end{align}
Since $\mathcal{A}^\dagger=\mathcal{A}^*$ and $\ket{\tilde{r}_m}=\ket{r_m}^*$, we have $\ket{\tilde{r}_{m+1}}=\ket{r_{m+1}}^*$.
By mathematical induction rule, the proof completes and we established between two subspaces $\ket{\tilde{r}_i}=(\ket{r_i})^*$ for all $i=0,1,2,\dots$. Under these conditions, the Eq.~\eqref{eq:bicg_orth} reduces to 
\begin{equation}\label{eq:orthogonality}
    \bra{r_j} r_i \rangle = \delta_{ij}
    \quad \text{ and  }\quad
    \bra{p_j} \mathcal{A}\ket{p_i}  = \delta_{ij}
    \,,
\end{equation}
which expresses the orthogonality among residuals and $\mathcal{A}$-orthogonality among search directions.
Therefore, in our specific scenario -- applicable to most physics problems-- the dual subspace $\mathcal{L}_m$ can be constructed directly from the original Krylov space without additional matrix-vector multiplications. This strategy effectively reduces the computational complexity of the original BiCG by $50$\%  bringing it on par with the cost of CG method. 

In this simplified setting, the usual two-term recurrence relations for the Krylov and dual residual updates in the BiCG method reduce to a single-term recurrence—resembling the structure of the CG algorithm. To be self-consist, we first establish the complete set of update equations for the seed system before extending them to the shifted systems in subsequent sections. For this purpose, we introduce an orthogonal search direction $\ket{p_m}$ along which the approximate solution vector is iteratively updated as
\begin{equation}\label{eq:seed_soln}
\ket{\Psi_{m+1}}=\ket{\Psi_m}+\alpha_m\ket{p_m}\,.
\end{equation}
Then the residual update follows
\begin{equation}\label{eq:seed_residual}
    \ket{r_{m+1}}= \ket{E}-\mathcal{A} \ket{\Psi_{m+1}}=\ket{r_m}-\alpha_m \mathcal{A}\ket{p_m}\,.
\end{equation}
We can find the value of scalar $\alpha_m$ by enforcing the orthogonality between residuals $\ket{r_m}$ and $\ket{r_{m+1}}$ i.e., $\bra{r_m}r_{m+1}\rangle=0$, which yields $\alpha_m=\bra{r_m}r_m\rangle/\bra{r_m}\mathcal{A}\ket{p_m}$. 
For serach direction update, we define
\begin{align}
    \ket{p_{m+1}} &= \ket{r_{m+1}} + \beta_{m} \ket{p_m}\,,
    \label{eq:seed_search_direction}
\end{align}
and require $\mathcal{A}$-orthogonality $\bra{p_m} \mathcal{A}\ket{p_{m+1}}=0$, which results $\beta_{m}=-\bra{r_{m+1}}\mathcal{A}\ket{p_m}/\bra{p_m}\mathcal{A}\ket{p_m}=\bra{r_{m+1}}r_{m+1}\rangle/\bra{r_m}r_m\rangle$. Although we solve these equations here for our specific case of real symmetric Hamiltonian, this method, in general, can be extended to any complex Hamiltonian by retaining the full Krylov and dual subspace structures along with conditions in Eq.~\eqref{eq:bicg_orth}.

\section{Brief Proof for the Identity of Krylov Subspaces and Colinearity of Residuals}\label{App:collinearity}
Here, we give a brief proof for the identity of the Krylov subspaces generated by shifted systems and the colinearity of the residuals generated using methods satisfying the (Petrov-)Galerkin condition.

Initial residuals of shifted systems Eq.~\eqref{eq:shifted_linear_eq} are given by
\begin{align}
    \ket{r_0^{(\win)}} &= \ket{E}-\left(\mathcal{A} - \win I\right)\ket{\Psi_0^{(\win)}}
   \label{eq:shift_init_residual}. 
\end{align}
If the initial guesses for both the seed and all shifted systems are set to zero, i.e.,  $\ket{\Psi_0}=\ket{0}$ and $\ket{\Psi_0^{(\win)}}=\ket{0}$, then all the initial residuals become identical and independent of the shift parameter, i.e., $\ket{r_0}=\ket{r_0^{(\win)}}=\ket{E}$ $\forall$ $\win$. In this case, it is straightforward that the Krylov space generated by the seed system and that of the shifted systems span the same $m$-dimensional subspace, i.e.,
\begin{equation}\label{eq:equal_Km}
    \mathcal{K}_m(\mathcal{A},\ket{r_0})=\mathcal{K}_m(\mathcal{A}-\win I,\ket{r_0}) \quad \text{for all $\win$.}
\end{equation}
This specific choice of initial guess ensures the collinearity among the initial residuals, i.e.,
\begin{equation}
    \ket{r_0^{(\win)}} = (1/\zeta_0^{(\win)})\ket{r_0},
\end{equation}
with $1/\zeta_0^{(\win)}=1$.
By combining Eq.~\eqref{eq:equal_Km} with (Petrov-)Galerkin condition, we can prove that the residuals across all shifted systems continue to remain collinear at each iteration step $m$. The (Petrov-)Galerkin condition guarantees that $\ket{r_m}\perp \mathcal{K}_m(\mathcal{A},\ket{r_0})$ and $\ket{r_m^{(\win)}}\perp \mathcal{K}_m(\mathcal{A}-\win I,\ket{r_0})$ (for $m=1,2,\dots$). From Eq.~\eqref{eq:equal_Km}, since the Krylov subspaces for the seed and shifted systems are identical, the residuals $\ket{r_m}$ and $\ket{r_m^{(\win)}}$ must lie in the same space and are collinear. Therefore, for some scalar $\zeta_m^{(\win)}\in \mathbb{C}$ 
\begin{equation}\label{eq:collinearity}
    \ket{r_m^{(\win)}} = (1/\zeta_m^{(\win)})\ket{r_m},
\end{equation}
which is the collinear equation discussed in Eq.~\eqref{eq:shiftCoeff}.

\section{Detailed Proof of the Iterative Equation}\label{App:Iterative_eq}
Following Refs.~\onlinecite{frommer2003BiCGStab, meng2015recycling}, we present the derivation of the MSBiCG algorithm, generalized here for resonant spectroscopy problems that involve solving families of shifted linear systems. As emphasized in the main text, the method exploits two key properties: the invariance of Krylov subspaces under shifts, and the collinearity of the associated residuals.

We noticed that solution $\ket{\Psi_m}\in \ket{\Psi_0}+ \mathcal{K}_m(\mathcal{A},\ket{r_0})$, where $\ket{r_0}=\ket{E}$, and residual $\ket{r_m}=\ket{E}-\mathcal{A}\ket{\Psi_m}\in\mathcal{K}_{m+1}(\mathcal{A},\ket{r_0})$. Expressing the solution as $\ket{\Psi_m}=\mathcal{Q}_{m-1}(\mathcal{A})\ket{r_0}$, where $\mathcal{Q}_{m-1}$ is a polynomial of degree at most $m-1$ with $\mathcal{Q}_{-1}(\mathcal{A})=0$, the $m^{\text{th}}$ residual becomes 
\begin{align}\label{eq:polynomial}
   \ket{r_m} &= \{I-\mathcal{A}\mathcal{Q}_{m-1}(\mathcal{A})\}\ket{r_0} =\mathcal{P}_m(\mathcal{A})\ket{r_0}, 
\end{align}
where $\mathcal{P}_m(\mathcal{A})=I-\mathcal{A}\mathcal{Q}_{m-1}(\mathcal{A})$ is the polynomial of degree at most $m$ with $\mathcal{P}_m(0)=1$. Substituting Eq.~\eqref{eq:polynomial} into Eq.~\eqref{eq:collinearity} and extending to all shifted systems gives
\begin{equation}\label{eq:poly_collinearity}
    \mathcal{P}_m^{(\win)}(\mathcal{A}-\win I) = (1/\zeta_m^{(\win)})\mathcal{P}_m(\mathcal{A})\,,
\end{equation}
a general identity that holds for arbitrary $\mathcal{A}$. Choosing $\mathcal{A}=\win I$ yields
\begin{equation}\label{eq:zeta}
    \zeta_m^{(\win)}=\mathcal{P}_m(\win)\,,
\end{equation}
which provides the fundamental link between the collinearity coefficient and the polynomial structure of the residuals. This relation forms the cornerstone for constructing the recurrence governing the shifted systems. 

To make this recurrence explicit, we rewrite Eqs.~\eqref{eq:seed_residual} and \eqref{eq:seed_search_direction} in the alternative forms:
\begin{align}
    \ket{p_{m-1}} &=\frac{1}{\beta_m}\left( \ket{p_m}-\ket{r_m} \right) 
    \label{eq:recast1}
    \\
    \mathcal{A}\ket{p_m} &=\frac{1}{\alpha_m}\left( \ket{r_m}-\ket{r_{m+1}}  \right)\,.
    \label{eq:recast2}
\end{align}
Substituting into the definition of BiCG residual $\ket{r_m}= \ket{r_{m-1}}-\alpha_{m-1}\mathcal{A}\ket{p_{m-1}}$ leads directly to the three-term residual recurrence relation for the seed system:
\begin{equation}\label{eq:seed_res_rec}
    \ket{r_{m+1}}=-\alpha_m \mathcal{A}\ket{r_m} + \left( 1+\frac{\alpha_m\beta_m}{\alpha_{m-1}} \right)\ket{r_m} - \frac{\alpha_m\beta_m}{\alpha_{m-1}}\ket{r_{m-1}}.
\end{equation}
The corresponding polynomial recurrence relation is obtained by substituting Eq.~\eqref{eq:polynomial} into Eq.~\eqref{eq:seed_res_rec}, which reads
\begin{widetext}
\begin{align}
\mathcal{P}_{m+1}(\mathcal{A}) &= 
-\alpha_m \mathcal{A}\mathcal{P}_m(\mathcal{A}) 
+ \left(1+\frac{\alpha_m \beta_{m}}{\alpha_{m-1}}\right)\mathcal{P}_m(\mathcal{A}) - \frac{\alpha_m \beta_{m}}{\alpha_{m-1}} \mathcal{P}_{m-1}(\mathcal{A})\,. 
\label{eq:poly_recurrsion}
\end{align}

Applying $\mathcal{A} = \win I$ and Eq.~\eqref{eq:zeta} in Eq.~\eqref{eq:poly_recurrsion}, we get the recurrence relation for collinear coefficients  
\begin{equation}\label{eq:zeta_recurr}
    \zeta_{m+1}^{(\win)} = \left(1+\alpha_m \win \right)\zeta_m^{(\win)} + \frac{\alpha_m\beta_{m}}{\alpha_{m-1}} \left( \zeta_m^{(\win)} - \zeta_{m-1}^{(\win)} \right),
\end{equation}
which reproduces Eq.~\eqref{eq:collinear_recursion} of the main text. Inserting $\ket{r_m}=\zeta_m^{(\win)}\ket{r_m^{(\win)}}$ into Eq.~\eqref{eq:seed_res_rec} and reorganizing terms leads to: 
\begin{align}\label{eq:shifted_recurssion1}
\ket{r_{m+1}^{(\win)}} &= 
-\alpha_m\left( \frac{\zeta_m^{(\win)}}{\zeta_{m+1}^{(\win)}} \right)  
(\mathcal{A}-\win I)\ket{r_m^{(\win)}} - \left( 1+\frac{\alpha_m \beta_{m}}{\alpha_{m-1}}
-\alpha_m\win \right)  
\frac{\zeta_m^{(\win)}}{\zeta_{m+1}^{(\win)}}  \ket{r_{m}^{(\win)}} - \frac{\alpha_m \beta_{m}}{\alpha_{m-1}} 
\frac{\zeta_{m-1}^{(\win)}}{\zeta_{m+1}^{(\win)}}  
\ket{r_{m-1}^{(\win)}}.
\end{align}
We can recast Eq.~\eqref{eq:shifted_recurssion1} into the form of seed residual recurrence in Eq.~\eqref{eq:seed_res_rec} as
\begin{align}\label{eq:shifted_recurssion}
\ket{r_{m+1}^{(\win)}} &=  
-\alpha_m^{(\win)}(\mathcal{A}-\win I)\ket{r_m^{(\win)}}  + \left( 1+\frac{\alpha_m^{(\win)} \beta_{m}^{(\win)}}{\alpha_{m-1}^{(\win)}} \right)\ket{r_{m}^{(\win)}}  - \frac{\alpha_m^{(\win)} \beta_{m}^{(\win)}}{\alpha_{m-1}^{(\win)}} \ket{r_{m-1}^{(\win)}}.
\end{align}
\end{widetext}

Recasting this in the BiCG three-term form gives
\begin{equation}
\alpha_m^{(\win)} = \left(\frac{\zeta_m^{(\win)}}{\zeta_{m+1}^{(\win)}}\right) \alpha_m \quad \text{and} \quad \beta_m^{(\win)} = \left(\frac{\zeta_{m-1}^{(\win)}}{\zeta_m^{(\win)}}\right)^2 \beta_m.
\label{eq:shift_coeffs}
\end{equation}
Eqs.~\eqref{eq:zeta_recurr} through \eqref{eq:shift_coeffs} hold for $m=0$ if we initialize $\zeta_{-1}^{(\win)}=1$. Now, following Eq.~\eqref{eq:seed_search_direction} and Eq.~\eqref{eq:seed_soln}, the corresponding updates for search direction and solution vector for multi-shift systems are generated by equations
\begin{align}
    \ket{p_m^{(\win)}} &= \left(1/ \zeta_m^{(\win)} \right)\ket{r_m} + \beta_m^{(\win)} \ket{p_{m-1}^{(\win)}}
\label{eq:search_dir_} \\
 \ket{\Psi_{m+1}^{(\win)}} &= \ket{\Psi_{m}^{(\win)}} + \alpha_m^{(\win)} \ket{p_m^{(\win)}}.
\label{eq:solution_ket_}
\end{align}
Together, the recursions in Eqs.~\eqref{eq:search_dir_} and \eqref{eq:solution_ket_} constitute the full multi-shifted BiCG method for constructing intermediate states in resonant spectroscopy.

\section{MSBiCG Algorithm}\label{App:MSBICG_algorithm}

\begin{algorithm}[H]
\caption{Multi-Shift BiCG for Resonant Spectroscopy}
\begin{algorithmic}[1]
\State Require: Matrix $\mathcal{A}$, excited state $\ket{E}$, shift frequencies $\{\win\}$, tolerance $\epsilon$, max iteration $M$ 
\State Initialize: $\ket{\Psi_0}=\ket{\Psi_0^{(\win)}}=\ket{0}$, $\ket{r_0}=\ket{r_0^{(\win)}}=\ket{E}$,  
 $\ket{p_0}=\ket{E}$, and $\ket{p_{-1}^{(\win)}}=\ket{0}$
 \State Initialize: $\zeta_{-1}^{(\win)}=1$,  $\zeta_{0}^{(\win)}=1$,  $\alpha_{-1}=1$
\For{$m = 0$ to $M$}
  \State $\ket{v_m}=\mathcal{A}\ket{p_m}$ \hspace{0.4 in}\(\textit{(matrix-vector multiplication)}\)
  \State 
  $\alpha_m={\langle r_m\ket{r_m}}/{\bra{r_m} v_m\rangle}$
  \State $\ket{\Psi_{m+1}}=\ket{\Psi_m}+\alpha_m\ket{p_m}$
  \State $\beta_{m}={\bra{r_{m+1}}r_{m+1}\rangle}/{\bra{r_{m}}r_{m}\rangle}$
  \State  $\ket{p_{m+1}}=\ket{r_{m+1}}+\beta_m \ket{p_{m}}$
  \State  $\ket{r_{m+1}} = \ket{r_m}-\alpha_m \ket{v_m}$
    
    \For{each active shift $\win$}
    \State $\begin{aligned}
   \zeta_{m+1}^{(\win)} = \left(1+\alpha_m \win \right)\zeta_m^{(\win)}   + \frac{\alpha_m\beta_{m}}{\alpha_{m-1}}
     \left(\zeta_m^{(\win)} - \zeta_{m-1}^{(\win)} \right)
\end{aligned}$
    \State $\alpha_m^{(\win)} = \left({\zeta_m^{(\win)}}/{\zeta_{m+1}^{(\win)}}\right) \alpha_m $
    \State $\beta_m^{(\win)} = \left({\zeta_{m-1}^{(\win)}}/{\zeta_m^{(\win)}}\right)^2 \beta_m$
    \State $  \ket{p_{m}^{(\win)}} = \left(1/ \zeta_m^{(\win)} \right)\ket{r_m} + \beta_m^{(\win)} \ket{p_{m-1}^{(\win)}}$
    \State $\ket{\Psi_{m+1}^{(\win)}} = \ket{\Psi_{m}^{(\win)}} + \alpha_m^{(\win)} \ket{p_m^{(\win)}}$
    \State $\ket{r_{m+1}^{(\win)}} = (1/\zeta_{m+1}^{(\win)})\ket{r_{m+1}}$

    \State Check convergence
    \EndFor

\EndFor
\end{algorithmic}
\end{algorithm}

\section{Adaptive Reseeding Strategy}\label{App:adaptive_reseeding}

Figure~\ref{fig:singleband_intensity}(c) in the main text illustrates that a fixed preselected seed can introduce substantial errors when multiple well-separated resonances contribute comparable intensities. The source of this problem lies in the accumulation of round-off errors within the recurrence Eq.~\eqref{eq:collinear_recursion}. While this relation is exact theoretically, numerical errors gradually build up over successive Krylov iterations, in a manner analogous to the loss of orthogonality in the Lanczos three-term recurrence. As a result, the collinearity condition is satisfied only approximately, and deviations in $\zeta_m^{(\win)}$ translate into errors in the shifted Lanczos coefficients $\alpha_m^{(\win)}$ and $\beta_m^{(\win)}$ through Eq.~\eqref{eq:coeffs}.

\begin{figure}[!b]
    \centering  
    \includegraphics[width=0.72\linewidth]{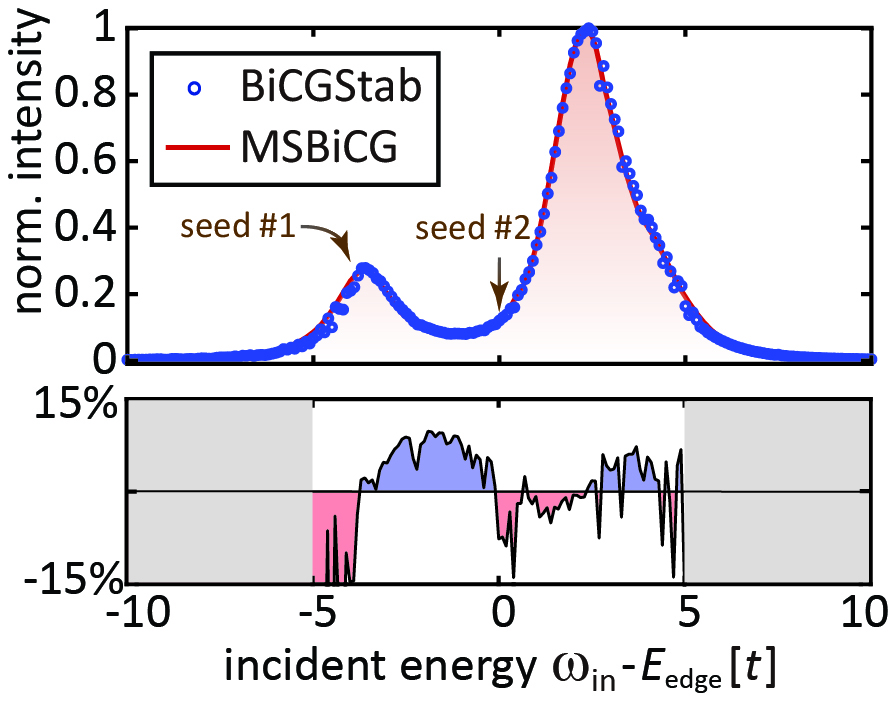}\vspace{-3mm}
    \caption{Upper: Comparison of BiCGStab and MSBiCG with the reseeding strategy applied along the $\omega = 5.47t$ cut of Fig.~\ref{fig:singleband_intensity}(c). The two arrows mark the incident energies $\win$ chosen as seed points. Lower: Relative spectral error evaluated near the resonance peaks along the same cut, plotted as a function of $\win$.}
    \label{fig:adaptive_reseeding}
\end{figure}

When the system is dominated by a single pole and the seed is chosen near that pole, these effects are negligible. In such cases, the shifted problems have smaller condition numbers than the seed problem, leading to rapid BiCG convergence and the irrelevance of the error accumulation. This is evident from the high accuracy of the spin-flip and spin-conserved direct RIXS simulations, which show negligible error [see Figs.~\ref{fig:singleband_intensity}(a) and (b)]. By contrast, when multiple resonances are widely separated and of comparable intensities, as in Fig.\ref{fig:singleband_intensity}(c), convergence is more demanding. The distant resonance requires many Krylov iterations to resolve, but in the MSBiCG framework its Krylov subspace is generated entirely from the recurrence relations in Eqs.~\eqref{eq:collinear_recursion} and \eqref{eq:coeffs}. The accumulated round-off error therefore plays a dominant role, degrading the accuracy and resulting in relative spectral errors of order $\sim 25\%$.

To address this issue and ensure the preservation of collinearity on the numerical level, we introduce an adaptive reseeding strategy. The procedure begins with an arbitrary seed selected within the incident-energy interval and updates it dynamically during the computation by monitoring deviations from residual collinearity. We define this deviation as $\Delta r^{(\win)}=\big\Vert\ket{r_t^{(\win)}}-\ket{r^{(\win)}}\big\Vert$, where $\ket{r_t^{(\win)}}=\ket{E}-(\mathcal{A}-\win I) \ket{\Psi^{(\win)}}$ is the true residual and $\ket{r^{(\win)}}$ is MSBiCG-simulated residual obtained from Eq.~\eqref{eq:collinearity}. Whenever $\Delta r^{(\win)}$ exceeds a prescribed tolerance, the corresponding $\win$ is promoted to serve as the new seed, and the iteration is restarted from this point. Although computing $\Delta r^{(\win)}$ requires a matrix–vector multiplication, this cost is incurred only once per $\win$, since the collinearity condition is independent of the iteration index $m$.  This adaptive reseeding strategy eliminates the need to predefine a seed (or incident energy) and ensures consistent accuracy across the spectrum. It also reduces the number of Krylov iterations by keeping each incident frequency close to its reference seed. 

We benchmark this strategy using the indirect RIXS calculation shown in Fig.~\ref{fig:singleband_intensity}(c) of the main text. With a fixed seed at $\win=0$, the earlier simulation exhibited a $23\%$ error at the resonance near $\win\sim-3.6t$. The adaptive reseeding results, shown in Fig.\ref{fig:adaptive_reseeding}, start with an initial seed at $\win=-4t$, which yields accurate results for $-10\leq\win<0$. Beyond this range, $\win=0$ is automatically promoted as the new seed for higher incident energies. This dynamic reseeding reduces the error at both resonance peaks to below $10\%$, demonstrating its effectiveness.

\bibliography{references_formatted}

\begin{thebibliography}{50}%
\makeatletter
\providecommand \@ifxundefined [1]{%
 \@ifx{#1\undefined}
}%
\providecommand \@ifnum [1]{%
 \ifnum #1\expandafter \@firstoftwo
 \else \expandafter \@secondoftwo
 \fi
}%
\providecommand \@ifx [1]{%
 \ifx #1\expandafter \@firstoftwo
 \else \expandafter \@secondoftwo
 \fi
}%
\providecommand \natexlab [1]{#1}%
\providecommand \enquote  [1]{``#1''}%
\providecommand \bibnamefont  [1]{#1}%
\providecommand \bibfnamefont [1]{#1}%
\providecommand \citenamefont [1]{#1}%
\providecommand \href@noop [0]{\@secondoftwo}%
\providecommand \href [0]{\begingroup \@sanitize@url \@href}%
\providecommand \@href[1]{\@@startlink{#1}\@@href}%
\providecommand \@@href[1]{\endgroup#1\@@endlink}%
\providecommand \@sanitize@url [0]{\catcode `\\12\catcode `\$12\catcode `\&12\catcode `\#12\catcode `\^12\catcode `\_12\catcode `\%12\relax}%
\providecommand \@@startlink[1]{}%
\providecommand \@@endlink[0]{}%
\providecommand \url  [0]{\begingroup\@sanitize@url \@url }%
\providecommand \@url [1]{\endgroup\@href {#1}{\urlprefix }}%
\providecommand \urlprefix  [0]{URL }%
\providecommand \Eprint [0]{\href }%
\providecommand \doibase [0]{https://doi.org/}%
\providecommand \selectlanguage [0]{\@gobble}%
\providecommand \bibinfo  [0]{\@secondoftwo}%
\providecommand \bibfield  [0]{\@secondoftwo}%
\providecommand \translation [1]{[#1]}%
\providecommand \BibitemOpen [0]{}%
\providecommand \bibitemStop [0]{}%
\providecommand \bibitemNoStop [0]{.\EOS\space}%
\providecommand \EOS [0]{\spacefactor3000\relax}%
\providecommand \BibitemShut  [1]{\csname bibitem#1\endcsname}%
\let\auto@bib@innerbib\@empty
\bibitem [{\citenamefont {Basov}\ \emph {et~al.}(2017)\citenamefont {Basov}, \citenamefont {Averitt},\ and\ \citenamefont {Hsieh}}]{basov2017towards}%
  \BibitemOpen
  \bibfield  {author} {\bibinfo {author} {\bibfnamefont {D.~N.}\ \bibnamefont {Basov}}, \bibinfo {author} {\bibfnamefont {R.~D.}\ \bibnamefont {Averitt}},\ and\ \bibinfo {author} {\bibfnamefont {D.}~\bibnamefont {Hsieh}},\ }\bibfield  {title} {\bibinfo {title} {{\textit{{Towards Properties on Demand in Quantum Materials}}}},\ }\href@noop {} {\bibfield  {journal} {\bibinfo  {journal} {Nat. Mater.}\ }\textbf {\bibinfo {volume} {16}},\ \bibinfo {pages} {1077} (\bibinfo {year} {2017})}\BibitemShut {NoStop}%
\bibitem [{\citenamefont {Keimer}\ and\ \citenamefont {Moore}(2017)}]{keimer2017physics}%
  \BibitemOpen
  \bibfield  {author} {\bibinfo {author} {\bibfnamefont {B.}~\bibnamefont {Keimer}}\ and\ \bibinfo {author} {\bibfnamefont {J.~E.}\ \bibnamefont {Moore}},\ }\bibfield  {title} {\bibinfo {title} {{\textit{{The Physics of Quantum Materials}}}},\ }\href@noop {} {\bibfield  {journal} {\bibinfo  {journal} {Nat. Phys.}\ }\textbf {\bibinfo {volume} {13}},\ \bibinfo {pages} {1045} (\bibinfo {year} {2017})}\BibitemShut {NoStop}%
\bibitem [{\citenamefont {de~Groot}\ \emph {et~al.}(2024)\citenamefont {de~Groot}, \citenamefont {Haverkort}, \citenamefont {Elnaggar}, \citenamefont {Juhin}, \citenamefont {Zhou},\ and\ \citenamefont {Glatzel}}]{de2024resonant}%
  \BibitemOpen
  \bibfield  {author} {\bibinfo {author} {\bibfnamefont {F.~M.}\ \bibnamefont {de~Groot}}, \bibinfo {author} {\bibfnamefont {M.~W.}\ \bibnamefont {Haverkort}}, \bibinfo {author} {\bibfnamefont {H.}~\bibnamefont {Elnaggar}}, \bibinfo {author} {\bibfnamefont {A.}~\bibnamefont {Juhin}}, \bibinfo {author} {\bibfnamefont {K.-J.}\ \bibnamefont {Zhou}},\ and\ \bibinfo {author} {\bibfnamefont {P.}~\bibnamefont {Glatzel}},\ }\bibfield  {title} {\bibinfo {title} {{\textit{Resonant Inelastic X-Ray Scattering}}},\ }\href@noop {} {\bibfield  {journal} {\bibinfo  {journal} {Nat. Rev. Methods Primers}\ }\textbf {\bibinfo {volume} {4}},\ \bibinfo {pages} {45} (\bibinfo {year} {2024})}\BibitemShut {NoStop}%
\bibitem [{\citenamefont {Mitrano}\ \emph {et~al.}(2024)\citenamefont {Mitrano}, \citenamefont {Johnston}, \citenamefont {Kim},\ and\ \citenamefont {Dean}}]{mitrano2024exploring}%
  \BibitemOpen
  \bibfield  {author} {\bibinfo {author} {\bibfnamefont {M.}~\bibnamefont {Mitrano}}, \bibinfo {author} {\bibfnamefont {S.}~\bibnamefont {Johnston}}, \bibinfo {author} {\bibfnamefont {Y.-J.}\ \bibnamefont {Kim}},\ and\ \bibinfo {author} {\bibfnamefont {M.}~\bibnamefont {Dean}},\ }\bibfield  {title} {\bibinfo {title} {{\textit{Exploring Quantum Materials with Resonant Inelastic X-Ray Scattering}}},\ }\href@noop {} {\bibfield  {journal} {\bibinfo  {journal} {Phys. Rev. X}\ }\textbf {\bibinfo {volume} {14}},\ \bibinfo {pages} {040501} (\bibinfo {year} {2024})}\BibitemShut {NoStop}%
\bibitem [{\citenamefont {Mitrano}\ and\ \citenamefont {Wang}(2020)}]{mitrano2020probing}%
  \BibitemOpen
  \bibfield  {author} {\bibinfo {author} {\bibfnamefont {M.}~\bibnamefont {Mitrano}}\ and\ \bibinfo {author} {\bibfnamefont {Y.}~\bibnamefont {Wang}},\ }\bibfield  {title} {\bibinfo {title} {{\textit{Probing Light-Driven Quantum Materials with Ultrafast Resonant Inelastic X-Ray Scattering}}},\ }\href@noop {} {\bibfield  {journal} {\bibinfo  {journal} {Commun. Phys.}\ }\textbf {\bibinfo {volume} {3}},\ \bibinfo {pages} {184} (\bibinfo {year} {2020})}\BibitemShut {NoStop}%
\bibitem [{\citenamefont {Liu}\ \emph {et~al.}(2025)\citenamefont {Liu}, \citenamefont {Xu}, \citenamefont {Liu},\ and\ \citenamefont {Wang}}]{liu2025entanglement}%
  \BibitemOpen
  \bibfield  {author} {\bibinfo {author} {\bibfnamefont {T.}~\bibnamefont {Liu}}, \bibinfo {author} {\bibfnamefont {L.}~\bibnamefont {Xu}}, \bibinfo {author} {\bibfnamefont {J.}~\bibnamefont {Liu}},\ and\ \bibinfo {author} {\bibfnamefont {Y.}~\bibnamefont {Wang}},\ }\bibfield  {title} {\bibinfo {title} {{\textit{Entanglement Witness for Indistinguishable Electrons Using Solid-State Spectroscopy}}},\ }\href@noop {} {\bibfield  {journal} {\bibinfo  {journal} {Phys. Rev. X}\ }\textbf {\bibinfo {volume} {15}},\ \bibinfo {pages} {011056} (\bibinfo {year} {2025})}\BibitemShut {NoStop}%
\bibitem [{\citenamefont {Kotani}\ and\ \citenamefont {Shin}(2001)}]{kotani2001resonant}%
  \BibitemOpen
  \bibfield  {author} {\bibinfo {author} {\bibfnamefont {A.}~\bibnamefont {Kotani}}\ and\ \bibinfo {author} {\bibfnamefont {S.}~\bibnamefont {Shin}},\ }\bibfield  {title} {\bibinfo {title} {{\textit{Resonant Inelastic X-Ray Scattering Spectra for Electrons in Solids}}},\ }\href@noop {} {\bibfield  {journal} {\bibinfo  {journal} {Rev. Mod. Phys.}\ }\textbf {\bibinfo {volume} {73}},\ \bibinfo {pages} {203} (\bibinfo {year} {2001})}\BibitemShut {NoStop}%
\bibitem [{\citenamefont {Ament}\ \emph {et~al.}(2011)\citenamefont {Ament}, \citenamefont {van Veenendaal}, \citenamefont {Devereaux}, \citenamefont {Hill},\ and\ \citenamefont {van~den Brink}}]{ament2011resonant}%
  \BibitemOpen
  \bibfield  {author} {\bibinfo {author} {\bibfnamefont {L.~J.~P.}\ \bibnamefont {Ament}}, \bibinfo {author} {\bibfnamefont {M.}~\bibnamefont {van Veenendaal}}, \bibinfo {author} {\bibfnamefont {T.~P.}\ \bibnamefont {Devereaux}}, \bibinfo {author} {\bibfnamefont {J.~P.}\ \bibnamefont {Hill}},\ and\ \bibinfo {author} {\bibfnamefont {J.}~\bibnamefont {van~den Brink}},\ }\bibfield  {title} {\bibinfo {title} {{\textit{Resonant Inelastic X-Ray Scattering Studies of Elementary Excitations}}},\ }\href@noop {} {\bibfield  {journal} {\bibinfo  {journal} {Rev. Mod. Phys.}\ }\textbf {\bibinfo {volume} {83}},\ \bibinfo {pages} {705} (\bibinfo {year} {2011})}\BibitemShut {NoStop}%
\bibitem [{\citenamefont {Ghiringhelli}\ \emph {et~al.}(2012)\citenamefont {Ghiringhelli}, \citenamefont {Le~Tacon}, \citenamefont {Minola}, \citenamefont {Blanco-Canosa}, \citenamefont {Mazzoli}, \citenamefont {Brookes}, \citenamefont {De~Luca}, \citenamefont {Frano}, \citenamefont {Hawthorn}, \citenamefont {He}, \citenamefont {Loew}, \citenamefont {Sala}, \citenamefont {Peets}, \citenamefont {Salluzzo}, \citenamefont {Schierle}, \citenamefont {Sutarto}, \citenamefont {Sawatzky}, \citenamefont {Weschke}, \citenamefont {Keimer},\ and\ \citenamefont {Braicovich}}]{ghiringhelli2012long}%
  \BibitemOpen
  \bibfield  {author} {\bibinfo {author} {\bibfnamefont {G.}~\bibnamefont {Ghiringhelli}}, \bibinfo {author} {\bibfnamefont {M.}~\bibnamefont {Le~Tacon}}, \bibinfo {author} {\bibfnamefont {M.}~\bibnamefont {Minola}}, \bibinfo {author} {\bibfnamefont {S.}~\bibnamefont {Blanco-Canosa}}, \bibinfo {author} {\bibfnamefont {C.}~\bibnamefont {Mazzoli}}, \bibinfo {author} {\bibfnamefont {N.}~\bibnamefont {Brookes}}, \bibinfo {author} {\bibfnamefont {G.}~\bibnamefont {De~Luca}}, \bibinfo {author} {\bibfnamefont {A.}~\bibnamefont {Frano}}, \bibinfo {author} {\bibfnamefont {D.}~\bibnamefont {Hawthorn}}, \bibinfo {author} {\bibfnamefont {F.}~\bibnamefont {He}}, \bibinfo {author} {\bibfnamefont {T.}~\bibnamefont {Loew}}, \bibinfo {author} {\bibfnamefont {M.~M.}\ \bibnamefont {Sala}}, \bibinfo {author} {\bibfnamefont {D.}~\bibnamefont {Peets}}, \bibinfo {author} {\bibfnamefont {M.}~\bibnamefont {Salluzzo}}, \bibinfo {author} {\bibfnamefont {E.}~\bibnamefont {Schierle}}, \bibinfo {author} {\bibfnamefont {R.}~\bibnamefont
  {Sutarto}}, \bibinfo {author} {\bibfnamefont {G.~A.}\ \bibnamefont {Sawatzky}}, \bibinfo {author} {\bibfnamefont {E.}~\bibnamefont {Weschke}}, \bibinfo {author} {\bibfnamefont {B.}~\bibnamefont {Keimer}},\ and\ \bibinfo {author} {\bibfnamefont {L.}~\bibnamefont {Braicovich}},\ }\bibfield  {title} {\bibinfo {title} {{\textit{{Long-Range Incommensurate Charge Fluctuations in (Y,Nd)Ba$_2$Cu$_3$O$_{6+x}$}}}},\ }\href@noop {} {\bibfield  {journal} {\bibinfo  {journal} {Science}\ }\textbf {\bibinfo {volume} {337}},\ \bibinfo {pages} {821} (\bibinfo {year} {2012})}\BibitemShut {NoStop}%
\bibitem [{\citenamefont {Dean}\ \emph {et~al.}(2013)\citenamefont {Dean}, \citenamefont {Dellea}, \citenamefont {Springell}, \citenamefont {Yakhou-Harris}, \citenamefont {Kummer}, \citenamefont {Brookes}, \citenamefont {Liu}, \citenamefont {Sun}, \citenamefont {Strle}, \citenamefont {Schmitt} \emph {et~al.}}]{dean2013persistence}%
  \BibitemOpen
  \bibfield  {author} {\bibinfo {author} {\bibfnamefont {M.}~\bibnamefont {Dean}}, \bibinfo {author} {\bibfnamefont {G.}~\bibnamefont {Dellea}}, \bibinfo {author} {\bibfnamefont {R.~S.}\ \bibnamefont {Springell}}, \bibinfo {author} {\bibfnamefont {F.}~\bibnamefont {Yakhou-Harris}}, \bibinfo {author} {\bibfnamefont {K.}~\bibnamefont {Kummer}}, \bibinfo {author} {\bibfnamefont {N.}~\bibnamefont {Brookes}}, \bibinfo {author} {\bibfnamefont {X.}~\bibnamefont {Liu}}, \bibinfo {author} {\bibfnamefont {Y.}~\bibnamefont {Sun}}, \bibinfo {author} {\bibfnamefont {J.}~\bibnamefont {Strle}}, \bibinfo {author} {\bibfnamefont {T.}~\bibnamefont {Schmitt}}, \emph {et~al.},\ }\bibfield  {title} {\bibinfo {title} {{\textit{Persistence of Magnetic Excitations in La$_{2-x}$Sr$_x$CuO$_4$ from the Undoped Insulator to the Heavily Overdoped Non-Superconducting Metal}}},\ }\href@noop {} {\bibfield  {journal} {\bibinfo  {journal} {Nat. Mater.}\ }\textbf {\bibinfo {volume} {12}},\ \bibinfo {pages} {1019} (\bibinfo {year}
  {2013})}\BibitemShut {NoStop}%
\bibitem [{\citenamefont {Hepting}\ \emph {et~al.}(2018)\citenamefont {Hepting}, \citenamefont {Chaix}, \citenamefont {Huang}, \citenamefont {Fumagalli}, \citenamefont {Peng}, \citenamefont {Moritz}, \citenamefont {Kummer}, \citenamefont {Brookes}, \citenamefont {Lee}, \citenamefont {Hashimoto} \emph {et~al.}}]{hepting2018three}%
  \BibitemOpen
  \bibfield  {author} {\bibinfo {author} {\bibfnamefont {M.}~\bibnamefont {Hepting}}, \bibinfo {author} {\bibfnamefont {L.}~\bibnamefont {Chaix}}, \bibinfo {author} {\bibfnamefont {E.}~\bibnamefont {Huang}}, \bibinfo {author} {\bibfnamefont {R.}~\bibnamefont {Fumagalli}}, \bibinfo {author} {\bibfnamefont {Y.}~\bibnamefont {Peng}}, \bibinfo {author} {\bibfnamefont {B.}~\bibnamefont {Moritz}}, \bibinfo {author} {\bibfnamefont {K.}~\bibnamefont {Kummer}}, \bibinfo {author} {\bibfnamefont {N.}~\bibnamefont {Brookes}}, \bibinfo {author} {\bibfnamefont {W.}~\bibnamefont {Lee}}, \bibinfo {author} {\bibfnamefont {M.}~\bibnamefont {Hashimoto}}, \emph {et~al.},\ }\bibfield  {title} {\bibinfo {title} {{\textit{Three-Dimensional Collective Charge Excitations in Electron-Doped Copper Oxide Superconductors}}},\ }\href@noop {} {\bibfield  {journal} {\bibinfo  {journal} {Nature}\ }\textbf {\bibinfo {volume} {563}},\ \bibinfo {pages} {374} (\bibinfo {year} {2018})}\BibitemShut {NoStop}%
\bibitem [{\citenamefont {Lu}\ \emph {et~al.}(2021)\citenamefont {Lu}, \citenamefont {Rossi}, \citenamefont {Nag}, \citenamefont {Osada}, \citenamefont {Li}, \citenamefont {Lee}, \citenamefont {Wang}, \citenamefont {Garcia-Fernandez}, \citenamefont {Agrestini}, \citenamefont {Shen}, \citenamefont {Been}, \citenamefont {Moritz}, \citenamefont {Devereaux}, \citenamefont {Zaanen}, \citenamefont {Hwang}, \citenamefont {Zhou},\ and\ \citenamefont {Lee}}]{lu2021magnetic}%
  \BibitemOpen
  \bibfield  {author} {\bibinfo {author} {\bibfnamefont {H.}~\bibnamefont {Lu}}, \bibinfo {author} {\bibfnamefont {M.}~\bibnamefont {Rossi}}, \bibinfo {author} {\bibfnamefont {A.}~\bibnamefont {Nag}}, \bibinfo {author} {\bibfnamefont {M.}~\bibnamefont {Osada}}, \bibinfo {author} {\bibfnamefont {D.}~\bibnamefont {Li}}, \bibinfo {author} {\bibfnamefont {K.}~\bibnamefont {Lee}}, \bibinfo {author} {\bibfnamefont {B.}~\bibnamefont {Wang}}, \bibinfo {author} {\bibfnamefont {M.}~\bibnamefont {Garcia-Fernandez}}, \bibinfo {author} {\bibfnamefont {S.}~\bibnamefont {Agrestini}}, \bibinfo {author} {\bibfnamefont {Z.~X.}\ \bibnamefont {Shen}}, \bibinfo {author} {\bibfnamefont {E.~M.}\ \bibnamefont {Been}}, \bibinfo {author} {\bibfnamefont {B.}~\bibnamefont {Moritz}}, \bibinfo {author} {\bibfnamefont {T.~P.}\ \bibnamefont {Devereaux}}, \bibinfo {author} {\bibfnamefont {J.}~\bibnamefont {Zaanen}}, \bibinfo {author} {\bibfnamefont {H.~Y.}\ \bibnamefont {Hwang}}, \bibinfo {author} {\bibfnamefont {K.-J.}\ \bibnamefont
  {Zhou}},\ and\ \bibinfo {author} {\bibfnamefont {W.~S.}\ \bibnamefont {Lee}},\ }\bibfield  {title} {\bibinfo {title} {{\textit{{Magnetic Excitations in Infinite-Layer Nickelates}}}},\ }\href@noop {} {\bibfield  {journal} {\bibinfo  {journal} {{Science}}\ }\textbf {\bibinfo {volume} {{373}}},\ \bibinfo {pages} {{213}} (\bibinfo {year} {{2021}})}\BibitemShut {NoStop}%
\bibitem [{\citenamefont {Gao}\ \emph {et~al.}(2024)\citenamefont {Gao}, \citenamefont {Fan}, \citenamefont {Wang}, \citenamefont {Li}, \citenamefont {Ren}, \citenamefont {Bia{\l}o}, \citenamefont {Drewanowski}, \citenamefont {Rothenb{\"u}hler}, \citenamefont {Choi}, \citenamefont {Sutarto} \emph {et~al.}}]{gao2024magnetic}%
  \BibitemOpen
  \bibfield  {author} {\bibinfo {author} {\bibfnamefont {Q.}~\bibnamefont {Gao}}, \bibinfo {author} {\bibfnamefont {S.}~\bibnamefont {Fan}}, \bibinfo {author} {\bibfnamefont {Q.}~\bibnamefont {Wang}}, \bibinfo {author} {\bibfnamefont {J.}~\bibnamefont {Li}}, \bibinfo {author} {\bibfnamefont {X.}~\bibnamefont {Ren}}, \bibinfo {author} {\bibfnamefont {I.}~\bibnamefont {Bia{\l}o}}, \bibinfo {author} {\bibfnamefont {A.}~\bibnamefont {Drewanowski}}, \bibinfo {author} {\bibfnamefont {P.}~\bibnamefont {Rothenb{\"u}hler}}, \bibinfo {author} {\bibfnamefont {J.}~\bibnamefont {Choi}}, \bibinfo {author} {\bibfnamefont {R.}~\bibnamefont {Sutarto}}, \emph {et~al.},\ }\bibfield  {title} {\bibinfo {title} {{\textit{Magnetic Excitations in Strained Infinite-Layer Nickelate PrNiO$_2$ Films}}},\ }\href@noop {} {\bibfield  {journal} {\bibinfo  {journal} {Nat. Commun.}\ }\textbf {\bibinfo {volume} {15}},\ \bibinfo {pages} {5576} (\bibinfo {year} {2024})}\BibitemShut {NoStop}%
\bibitem [{\citenamefont {Chen}\ \emph {et~al.}(2024)\citenamefont {Chen}, \citenamefont {Choi}, \citenamefont {Jiang}, \citenamefont {Mei}, \citenamefont {Jiang}, \citenamefont {Li}, \citenamefont {Agrestini}, \citenamefont {Garcia-Fernandez}, \citenamefont {Sun}, \citenamefont {Huang} \emph {et~al.}}]{chen2024electronic}%
  \BibitemOpen
  \bibfield  {author} {\bibinfo {author} {\bibfnamefont {X.}~\bibnamefont {Chen}}, \bibinfo {author} {\bibfnamefont {J.}~\bibnamefont {Choi}}, \bibinfo {author} {\bibfnamefont {Z.}~\bibnamefont {Jiang}}, \bibinfo {author} {\bibfnamefont {J.}~\bibnamefont {Mei}}, \bibinfo {author} {\bibfnamefont {K.}~\bibnamefont {Jiang}}, \bibinfo {author} {\bibfnamefont {J.}~\bibnamefont {Li}}, \bibinfo {author} {\bibfnamefont {S.}~\bibnamefont {Agrestini}}, \bibinfo {author} {\bibfnamefont {M.}~\bibnamefont {Garcia-Fernandez}}, \bibinfo {author} {\bibfnamefont {H.}~\bibnamefont {Sun}}, \bibinfo {author} {\bibfnamefont {X.}~\bibnamefont {Huang}}, \emph {et~al.},\ }\bibfield  {title} {\bibinfo {title} {{\textit{Electronic and Magnetic Excitations in La$_3$Ni$_2$O$_7$}}},\ }\href@noop {} {\bibfield  {journal} {\bibinfo  {journal} {Nat. Commun.}\ }\textbf {\bibinfo {volume} {15}},\ \bibinfo {pages} {9597} (\bibinfo {year} {2024})}\BibitemShut {NoStop}%
\bibitem [{\citenamefont {Kim}\ \emph {et~al.}(2012)\citenamefont {Kim}, \citenamefont {Casa}, \citenamefont {Upton}, \citenamefont {Gog}, \citenamefont {Kim}, \citenamefont {Mitchell}, \citenamefont {Van~Veenendaal}, \citenamefont {Daghofer}, \citenamefont {Van Den~Brink}, \citenamefont {Khaliullin} \emph {et~al.}}]{kim2012magnetic}%
  \BibitemOpen
  \bibfield  {author} {\bibinfo {author} {\bibfnamefont {J.}~\bibnamefont {Kim}}, \bibinfo {author} {\bibfnamefont {D.}~\bibnamefont {Casa}}, \bibinfo {author} {\bibfnamefont {M.}~\bibnamefont {Upton}}, \bibinfo {author} {\bibfnamefont {T.}~\bibnamefont {Gog}}, \bibinfo {author} {\bibfnamefont {Y.-J.}\ \bibnamefont {Kim}}, \bibinfo {author} {\bibfnamefont {J.}~\bibnamefont {Mitchell}}, \bibinfo {author} {\bibfnamefont {M.}~\bibnamefont {Van~Veenendaal}}, \bibinfo {author} {\bibfnamefont {M.}~\bibnamefont {Daghofer}}, \bibinfo {author} {\bibfnamefont {J.}~\bibnamefont {Van Den~Brink}}, \bibinfo {author} {\bibfnamefont {G.}~\bibnamefont {Khaliullin}}, \emph {et~al.},\ }\bibfield  {title} {\bibinfo {title} {{\textit{Magnetic Excitation Spectra of Sr$_2$IrO$_4$ Probed by Resonant Inelastic X-Ray Scattering: Establishing Links to Cuprate Superconductors}}},\ }\href@noop {} {\bibfield  {journal} {\bibinfo  {journal} {Phys. Rev. Lett.}\ }\textbf {\bibinfo {volume} {108}},\ \bibinfo {pages} {177003} (\bibinfo {year}
  {2012})}\BibitemShut {NoStop}%
\bibitem [{\citenamefont {Pelliciari}\ \emph {et~al.}(2021)\citenamefont {Pelliciari}, \citenamefont {Karakuzu}, \citenamefont {Song}, \citenamefont {Arpaia}, \citenamefont {Nag}, \citenamefont {Rossi}, \citenamefont {Li}, \citenamefont {Yu}, \citenamefont {Chen}, \citenamefont {Peng} \emph {et~al.}}]{pelliciari2021evolution}%
  \BibitemOpen
  \bibfield  {author} {\bibinfo {author} {\bibfnamefont {J.}~\bibnamefont {Pelliciari}}, \bibinfo {author} {\bibfnamefont {S.}~\bibnamefont {Karakuzu}}, \bibinfo {author} {\bibfnamefont {Q.}~\bibnamefont {Song}}, \bibinfo {author} {\bibfnamefont {R.}~\bibnamefont {Arpaia}}, \bibinfo {author} {\bibfnamefont {A.}~\bibnamefont {Nag}}, \bibinfo {author} {\bibfnamefont {M.}~\bibnamefont {Rossi}}, \bibinfo {author} {\bibfnamefont {J.}~\bibnamefont {Li}}, \bibinfo {author} {\bibfnamefont {T.}~\bibnamefont {Yu}}, \bibinfo {author} {\bibfnamefont {X.}~\bibnamefont {Chen}}, \bibinfo {author} {\bibfnamefont {R.}~\bibnamefont {Peng}}, \emph {et~al.},\ }\bibfield  {title} {\bibinfo {title} {{\textit{Evolution of Spin Excitations from Bulk to Monolayer FeSe}}},\ }\href@noop {} {\bibfield  {journal} {\bibinfo  {journal} {Nat. Commun.}\ }\textbf {\bibinfo {volume} {12}},\ \bibinfo {pages} {3122} (\bibinfo {year} {2021})}\BibitemShut {NoStop}%
\bibitem [{\citenamefont {Zhang}\ \emph {et~al.}(2019)\citenamefont {Zhang}, \citenamefont {Li}, \citenamefont {Ouyang}, \citenamefont {Yu}, \citenamefont {Ge}, \citenamefont {Huang}, \citenamefont {Hu}, \citenamefont {Ma}, \citenamefont {Li}, \citenamefont {Xiao} \emph {et~al.}}]{zhang2019trace}%
  \BibitemOpen
  \bibfield  {author} {\bibinfo {author} {\bibfnamefont {J.-N.}\ \bibnamefont {Zhang}}, \bibinfo {author} {\bibfnamefont {Q.}~\bibnamefont {Li}}, \bibinfo {author} {\bibfnamefont {C.}~\bibnamefont {Ouyang}}, \bibinfo {author} {\bibfnamefont {X.}~\bibnamefont {Yu}}, \bibinfo {author} {\bibfnamefont {M.}~\bibnamefont {Ge}}, \bibinfo {author} {\bibfnamefont {X.}~\bibnamefont {Huang}}, \bibinfo {author} {\bibfnamefont {E.}~\bibnamefont {Hu}}, \bibinfo {author} {\bibfnamefont {C.}~\bibnamefont {Ma}}, \bibinfo {author} {\bibfnamefont {S.}~\bibnamefont {Li}}, \bibinfo {author} {\bibfnamefont {R.}~\bibnamefont {Xiao}}, \emph {et~al.},\ }\bibfield  {title} {\bibinfo {title} {{\textit{Trace Doping of Multiple Elements Enables Stable Battery Cycling of LiCoO$_2$ at 4.6\,V}}},\ }\href@noop {} {\bibfield  {journal} {\bibinfo  {journal} {Nature Energy}\ }\textbf {\bibinfo {volume} {4}},\ \bibinfo {pages} {594} (\bibinfo {year} {2019})}\BibitemShut {NoStop}%
\bibitem [{\citenamefont {Mahmood}\ \emph {et~al.}(2022)\citenamefont {Mahmood}, \citenamefont {Devereaux}, \citenamefont {Abbamonte},\ and\ \citenamefont {Morr}}]{mahmood2022distinguishing}%
  \BibitemOpen
  \bibfield  {author} {\bibinfo {author} {\bibfnamefont {F.}~\bibnamefont {Mahmood}}, \bibinfo {author} {\bibfnamefont {T.}~\bibnamefont {Devereaux}}, \bibinfo {author} {\bibfnamefont {P.}~\bibnamefont {Abbamonte}},\ and\ \bibinfo {author} {\bibfnamefont {D.~K.}\ \bibnamefont {Morr}},\ }\bibfield  {title} {\bibinfo {title} {{\textit{Distinguishing Finite-Momentum Superconducting Pairing States with Two-Electron Photoemission Spectroscopy}}},\ }\href@noop {} {\bibfield  {journal} {\bibinfo  {journal} {Phys. Rev. B}\ }\textbf {\bibinfo {volume} {105}},\ \bibinfo {pages} {064515} (\bibinfo {year} {2022})}\BibitemShut {NoStop}%
\bibitem [{\citenamefont {Peticolas}\ \emph {et~al.}(1963)\citenamefont {Peticolas}, \citenamefont {Goldsborough},\ and\ \citenamefont {Rieckhoff}}]{peticolas1963double}%
  \BibitemOpen
  \bibfield  {author} {\bibinfo {author} {\bibfnamefont {W.~L.}\ \bibnamefont {Peticolas}}, \bibinfo {author} {\bibfnamefont {J.~P.}\ \bibnamefont {Goldsborough}},\ and\ \bibinfo {author} {\bibfnamefont {K.}~\bibnamefont {Rieckhoff}},\ }\bibfield  {title} {\bibinfo {title} {{\textit{Double Photon Excitation in Organic Crystals}}},\ }\href@noop {} {\bibfield  {journal} {\bibinfo  {journal} {Phys. Rev. Lett.}\ }\textbf {\bibinfo {volume} {10}},\ \bibinfo {pages} {43} (\bibinfo {year} {1963})}\BibitemShut {NoStop}%
\bibitem [{\citenamefont {Devereaux}\ and\ \citenamefont {Hackl}(2007)}]{devereaux2007inelastic}%
  \BibitemOpen
  \bibfield  {author} {\bibinfo {author} {\bibfnamefont {T.~P.}\ \bibnamefont {Devereaux}}\ and\ \bibinfo {author} {\bibfnamefont {R.}~\bibnamefont {Hackl}},\ }\bibfield  {title} {\bibinfo {title} {{\textit{Inelastic Light Scattering from Correlated Electrons}}},\ }\href@noop {} {\bibfield  {journal} {\bibinfo  {journal} {Rev. Mod. Phys.}\ }\textbf {\bibinfo {volume} {79}},\ \bibinfo {pages} {175–233} (\bibinfo {year} {2007})}\BibitemShut {NoStop}%
\bibitem [{\citenamefont {Ko}\ \emph {et~al.}(2010)\citenamefont {Ko}, \citenamefont {Liu}, \citenamefont {Ng},\ and\ \citenamefont {Lee}}]{ko2010raman}%
  \BibitemOpen
  \bibfield  {author} {\bibinfo {author} {\bibfnamefont {W.-H.}\ \bibnamefont {Ko}}, \bibinfo {author} {\bibfnamefont {Z.-X.}\ \bibnamefont {Liu}}, \bibinfo {author} {\bibfnamefont {T.-K.}\ \bibnamefont {Ng}},\ and\ \bibinfo {author} {\bibfnamefont {P.~A.}\ \bibnamefont {Lee}},\ }\bibfield  {title} {\bibinfo {title} {{\textit{{Raman Signature of the U(1) Dirac Spin-Liquid State in the Spin-1/2 Kagome System}}}},\ }\href@noop {} {\bibfield  {journal} {\bibinfo  {journal} {Phys. Rev. B}\ }\textbf {\bibinfo {volume} {81}},\ \bibinfo {pages} {024414} (\bibinfo {year} {2010})}\BibitemShut {NoStop}%
\bibitem [{\citenamefont {de~Farias}\ \emph {et~al.}(2020)\citenamefont {de~Farias}, \citenamefont {M{\'e}asson}, \citenamefont {Ferraz},\ and\ \citenamefont {Burdin}}]{de2020effective}%
  \BibitemOpen
  \bibfield  {author} {\bibinfo {author} {\bibfnamefont {C.~S.}\ \bibnamefont {de~Farias}}, \bibinfo {author} {\bibfnamefont {M.-A.}\ \bibnamefont {M{\'e}asson}}, \bibinfo {author} {\bibfnamefont {A.}~\bibnamefont {Ferraz}},\ and\ \bibinfo {author} {\bibfnamefont {S.}~\bibnamefont {Burdin}},\ }\bibfield  {title} {\bibinfo {title} {{\textit{{Effective Model for the $A_{2g}$ Raman Signal in UrU$_2$Si$_2$}}}},\ }\href@noop {} {\bibfield  {journal} {\bibinfo  {journal} {Phys. Rev. B}\ }\textbf {\bibinfo {volume} {101}},\ \bibinfo {pages} {205114} (\bibinfo {year} {2020})}\BibitemShut {NoStop}%
\bibitem [{\citenamefont {Devereaux}\ \emph {et~al.}(2023)\citenamefont {Devereaux}, \citenamefont {Claassen}, \citenamefont {Huang}, \citenamefont {Zaletel}, \citenamefont {Moore}, \citenamefont {Morr}, \citenamefont {Mahmood}, \citenamefont {Abbamonte},\ and\ \citenamefont {Shen}}]{devereaux2023angle}%
  \BibitemOpen
  \bibfield  {author} {\bibinfo {author} {\bibfnamefont {T.~P.}\ \bibnamefont {Devereaux}}, \bibinfo {author} {\bibfnamefont {M.}~\bibnamefont {Claassen}}, \bibinfo {author} {\bibfnamefont {X.-X.}\ \bibnamefont {Huang}}, \bibinfo {author} {\bibfnamefont {M.}~\bibnamefont {Zaletel}}, \bibinfo {author} {\bibfnamefont {J.~E.}\ \bibnamefont {Moore}}, \bibinfo {author} {\bibfnamefont {D.}~\bibnamefont {Morr}}, \bibinfo {author} {\bibfnamefont {F.}~\bibnamefont {Mahmood}}, \bibinfo {author} {\bibfnamefont {P.}~\bibnamefont {Abbamonte}},\ and\ \bibinfo {author} {\bibfnamefont {Z.-X.}\ \bibnamefont {Shen}},\ }\bibfield  {title} {\bibinfo {title} {{\textit{Angle-Resolved Pair Photoemission Theory for Correlated Electrons}}},\ }\href@noop {} {\bibfield  {journal} {\bibinfo  {journal} {Phys. Rev. B}\ }\textbf {\bibinfo {volume} {108}},\ \bibinfo {pages} {165134} (\bibinfo {year} {2023})}\BibitemShut {NoStop}%
\bibitem [{\citenamefont {Hsu}\ \emph {et~al.}(2025)\citenamefont {Hsu}, \citenamefont {Jia}, \citenamefont {Zhang}, \citenamefont {Jost}, \citenamefont {Moritz}, \citenamefont {Hackl},\ and\ \citenamefont {Devereaux}}]{hsu2025detection}%
  \BibitemOpen
  \bibfield  {author} {\bibinfo {author} {\bibfnamefont {K.~H.}\ \bibnamefont {Hsu}}, \bibinfo {author} {\bibfnamefont {C.}~\bibnamefont {Jia}}, \bibinfo {author} {\bibfnamefont {E.~Z.}\ \bibnamefont {Zhang}}, \bibinfo {author} {\bibfnamefont {D.}~\bibnamefont {Jost}}, \bibinfo {author} {\bibfnamefont {B.}~\bibnamefont {Moritz}}, \bibinfo {author} {\bibfnamefont {R.}~\bibnamefont {Hackl}},\ and\ \bibinfo {author} {\bibfnamefont {T.~P.}\ \bibnamefont {Devereaux}},\ }\bibfield  {title} {\bibinfo {title} {{\textit{Detection of Chiral Spin Fluctuations Driven by Frustration in Mott Insulators}}},\ }\href@noop {} {\bibfield  {journal} {\bibinfo  {journal} {Phys. Rev. B}\ }\textbf {\bibinfo {volume} {111}},\ \bibinfo {pages} {205115} (\bibinfo {year} {2025})}\BibitemShut {NoStop}%
\bibitem [{\citenamefont {Wang}\ \emph {et~al.}(2018)\citenamefont {Wang}, \citenamefont {Claassen}, \citenamefont {Pemmaraju}, \citenamefont {Jia}, \citenamefont {Moritz},\ and\ \citenamefont {Devereaux}}]{wang2018theoretical}%
  \BibitemOpen
  \bibfield  {author} {\bibinfo {author} {\bibfnamefont {Y.}~\bibnamefont {Wang}}, \bibinfo {author} {\bibfnamefont {M.}~\bibnamefont {Claassen}}, \bibinfo {author} {\bibfnamefont {C.~D.}\ \bibnamefont {Pemmaraju}}, \bibinfo {author} {\bibfnamefont {C.}~\bibnamefont {Jia}}, \bibinfo {author} {\bibfnamefont {B.}~\bibnamefont {Moritz}},\ and\ \bibinfo {author} {\bibfnamefont {T.~P.}\ \bibnamefont {Devereaux}},\ }\bibfield  {title} {\bibinfo {title} {{\textit{Theoretical Understanding of Photon Spectroscopies in Correlated Materials in and out of Equilibrium}}},\ }\href@noop {} {\bibfield  {journal} {\bibinfo  {journal} {Nat. Rev. Mater.}\ }\textbf {\bibinfo {volume} {3}},\ \bibinfo {pages} {312} (\bibinfo {year} {2018})}\BibitemShut {NoStop}%
\bibitem [{\citenamefont {Ament}\ \emph {et~al.}(2009)\citenamefont {Ament}, \citenamefont {Ghiringhelli}, \citenamefont {Sala}, \citenamefont {Braicovich},\ and\ \citenamefont {van~den Brink}}]{ament2009theoretical}%
  \BibitemOpen
  \bibfield  {author} {\bibinfo {author} {\bibfnamefont {L.~J.}\ \bibnamefont {Ament}}, \bibinfo {author} {\bibfnamefont {G.}~\bibnamefont {Ghiringhelli}}, \bibinfo {author} {\bibfnamefont {M.~M.}\ \bibnamefont {Sala}}, \bibinfo {author} {\bibfnamefont {L.}~\bibnamefont {Braicovich}},\ and\ \bibinfo {author} {\bibfnamefont {J.}~\bibnamefont {van~den Brink}},\ }\bibfield  {title} {\bibinfo {title} {{\textit{Theoretical Demonstration of How the Dispersion of Magnetic Excitations in Cuprate Compounds Can Be Determined Using Resonant Inelastic X-Ray Scattering}}},\ }\href@noop {} {\bibfield  {journal} {\bibinfo  {journal} {Phys. Rev. Lett.}\ }\textbf {\bibinfo {volume} {103}},\ \bibinfo {pages} {117003} (\bibinfo {year} {2009})}\BibitemShut {NoStop}%
\bibitem [{\citenamefont {van~den Brink}(2007)}]{van2007theory}%
  \BibitemOpen
  \bibfield  {author} {\bibinfo {author} {\bibfnamefont {J.}~\bibnamefont {van~den Brink}},\ }\bibfield  {title} {\bibinfo {title} {{\textit{The Theory of Indirect Resonant Inelastic X-Ray Scattering on Magnons}}},\ }\href@noop {} {\bibfield  {journal} {\bibinfo  {journal} {Europhys. Lett.}\ }\textbf {\bibinfo {volume} {80}},\ \bibinfo {pages} {47003} (\bibinfo {year} {2007})}\BibitemShut {NoStop}%
\bibitem [{\citenamefont {Braicovich}\ \emph {et~al.}(2009)\citenamefont {Braicovich}, \citenamefont {Ament}, \citenamefont {Bisogni}, \citenamefont {Forte}, \citenamefont {Aruta}, \citenamefont {Balestrino}, \citenamefont {Brookes}, \citenamefont {De~Luca}, \citenamefont {Medaglia}, \citenamefont {Granozio} \emph {et~al.}}]{braicovich2009dispersion}%
  \BibitemOpen
  \bibfield  {author} {\bibinfo {author} {\bibfnamefont {L.}~\bibnamefont {Braicovich}}, \bibinfo {author} {\bibfnamefont {L.}~\bibnamefont {Ament}}, \bibinfo {author} {\bibfnamefont {V.}~\bibnamefont {Bisogni}}, \bibinfo {author} {\bibfnamefont {F.}~\bibnamefont {Forte}}, \bibinfo {author} {\bibfnamefont {C.}~\bibnamefont {Aruta}}, \bibinfo {author} {\bibfnamefont {G.}~\bibnamefont {Balestrino}}, \bibinfo {author} {\bibfnamefont {N.}~\bibnamefont {Brookes}}, \bibinfo {author} {\bibfnamefont {G.}~\bibnamefont {De~Luca}}, \bibinfo {author} {\bibfnamefont {P.}~\bibnamefont {Medaglia}}, \bibinfo {author} {\bibfnamefont {F.~M.}\ \bibnamefont {Granozio}}, \emph {et~al.},\ }\bibfield  {title} {\bibinfo {title} {{\textit{Dispersion of Magnetic Excitations in the Cuprate La$_2$CuO$_4$ and CaCuO$_2$ Compounds Measured Using Resonant X-Ray Scattering}}},\ }\href@noop {} {\bibfield  {journal} {\bibinfo  {journal} {Phys. Rev. Lett.}\ }\textbf {\bibinfo {volume} {102}},\ \bibinfo {pages} {167401} (\bibinfo {year}
  {2009})}\BibitemShut {NoStop}%
\bibitem [{\citenamefont {Forte}\ \emph {et~al.}(2008)\citenamefont {Forte}, \citenamefont {Ament},\ and\ \citenamefont {van~den Brink}}]{forte2008magnetic}%
  \BibitemOpen
  \bibfield  {author} {\bibinfo {author} {\bibfnamefont {F.}~\bibnamefont {Forte}}, \bibinfo {author} {\bibfnamefont {L.~J.}\ \bibnamefont {Ament}},\ and\ \bibinfo {author} {\bibfnamefont {J.}~\bibnamefont {van~den Brink}},\ }\bibfield  {title} {\bibinfo {title} {{\textit{Magnetic Excitations in La$_2$CuO$_4$ Probed by Indirect Resonant Inelastic X-Ray Scattering}}},\ }\href@noop {} {\bibfield  {journal} {\bibinfo  {journal} {Phys. Rev. B}\ }\textbf {\bibinfo {volume} {77}},\ \bibinfo {pages} {134428} (\bibinfo {year} {2008})}\BibitemShut {NoStop}%
\bibitem [{\citenamefont {Dagotto}(1994)}]{dagotto1994correlated}%
  \BibitemOpen
  \bibfield  {author} {\bibinfo {author} {\bibfnamefont {E.}~\bibnamefont {Dagotto}},\ }\bibfield  {title} {\bibinfo {title} {{\textit{Correlated Electrons in High-Temperature Superconductors}}},\ }\href@noop {} {\bibfield  {journal} {\bibinfo  {journal} {Rev. Mod. Phys.}\ }\textbf {\bibinfo {volume} {66}},\ \bibinfo {pages} {763} (\bibinfo {year} {1994})}\BibitemShut {NoStop}%
\bibitem [{\citenamefont {Jia}\ \emph {et~al.}(2016)\citenamefont {Jia}, \citenamefont {Wohlfeld}, \citenamefont {Wang}, \citenamefont {Moritz},\ and\ \citenamefont {Devereaux}}]{jia2016using}%
  \BibitemOpen
  \bibfield  {author} {\bibinfo {author} {\bibfnamefont {C.}~\bibnamefont {Jia}}, \bibinfo {author} {\bibfnamefont {K.}~\bibnamefont {Wohlfeld}}, \bibinfo {author} {\bibfnamefont {Y.}~\bibnamefont {Wang}}, \bibinfo {author} {\bibfnamefont {B.}~\bibnamefont {Moritz}},\ and\ \bibinfo {author} {\bibfnamefont {T.~P.}\ \bibnamefont {Devereaux}},\ }\bibfield  {title} {\bibinfo {title} {{\textit{Using RIXS to Uncover Elementary Charge and Spin Excitations}}},\ }\href@noop {} {\bibfield  {journal} {\bibinfo  {journal} {Phys. Rev. X}\ }\textbf {\bibinfo {volume} {6}},\ \bibinfo {pages} {021020} (\bibinfo {year} {2016})}\BibitemShut {NoStop}%
\bibitem [{\citenamefont {Wang}\ \emph {et~al.}(2019)\citenamefont {Wang}, \citenamefont {Fabbris}, \citenamefont {Dean},\ and\ \citenamefont {Kotliar}}]{wang2019edrixs}%
  \BibitemOpen
  \bibfield  {author} {\bibinfo {author} {\bibfnamefont {Y.}~\bibnamefont {Wang}}, \bibinfo {author} {\bibfnamefont {G.}~\bibnamefont {Fabbris}}, \bibinfo {author} {\bibfnamefont {M.~P.}\ \bibnamefont {Dean}},\ and\ \bibinfo {author} {\bibfnamefont {G.}~\bibnamefont {Kotliar}},\ }\bibfield  {title} {\bibinfo {title} {{\textit{EDRIXS: An Open Source Toolkit for Simulating Spectra of Resonant Inelastic X-Ray Scattering}}},\ }\href@noop {} {\bibfield  {journal} {\bibinfo  {journal} {Comput. Phys. Commun.}\ }\textbf {\bibinfo {volume} {243}},\ \bibinfo {pages} {151} (\bibinfo {year} {2019})}\BibitemShut {NoStop}%
\bibitem [{\citenamefont {Chen}\ \emph {et~al.}(2010)\citenamefont {Chen}, \citenamefont {Moritz}, \citenamefont {Vernay}, \citenamefont {Hancock}, \citenamefont {Johnston}, \citenamefont {Jia}, \citenamefont {Chabot-Couture}, \citenamefont {Greven}, \citenamefont {Elfimov}, \citenamefont {Sawatzky} \emph {et~al.}}]{chen2010unraveling}%
  \BibitemOpen
  \bibfield  {author} {\bibinfo {author} {\bibfnamefont {C.-C.}\ \bibnamefont {Chen}}, \bibinfo {author} {\bibfnamefont {B.}~\bibnamefont {Moritz}}, \bibinfo {author} {\bibfnamefont {F.}~\bibnamefont {Vernay}}, \bibinfo {author} {\bibfnamefont {J.}~\bibnamefont {Hancock}}, \bibinfo {author} {\bibfnamefont {S.}~\bibnamefont {Johnston}}, \bibinfo {author} {\bibfnamefont {C.}~\bibnamefont {Jia}}, \bibinfo {author} {\bibfnamefont {G.}~\bibnamefont {Chabot-Couture}}, \bibinfo {author} {\bibfnamefont {M.}~\bibnamefont {Greven}}, \bibinfo {author} {\bibfnamefont {I.}~\bibnamefont {Elfimov}}, \bibinfo {author} {\bibfnamefont {G.}~\bibnamefont {Sawatzky}}, \emph {et~al.},\ }\bibfield  {title} {\bibinfo {title} {\textit{Unraveling the Nature of Charge Excitations in La$_2$CuO$_4$ with Momentum-Resolved Cu K-Edge Resonant Inelastic X-Ray Scattering}},\ }\href@noop {} {\bibfield  {journal} {\bibinfo  {journal} {Phys. Rev. Lett.}\ }\textbf {\bibinfo {volume} {105}},\ \bibinfo {pages} {177401} (\bibinfo {year}
  {2010})}\BibitemShut {NoStop}%
\bibitem [{\citenamefont {Jia}\ \emph {et~al.}(2012)\citenamefont {Jia}, \citenamefont {Chen}, \citenamefont {Sorini}, \citenamefont {Moritz},\ and\ \citenamefont {Devereaux}}]{jia2012uncovering}%
  \BibitemOpen
  \bibfield  {author} {\bibinfo {author} {\bibfnamefont {C.}~\bibnamefont {Jia}}, \bibinfo {author} {\bibfnamefont {C.}~\bibnamefont {Chen}}, \bibinfo {author} {\bibfnamefont {A.}~\bibnamefont {Sorini}}, \bibinfo {author} {\bibfnamefont {B.}~\bibnamefont {Moritz}},\ and\ \bibinfo {author} {\bibfnamefont {T.}~\bibnamefont {Devereaux}},\ }\bibfield  {title} {\bibinfo {title} {{\textit{Uncovering Selective Excitations Using the Resonant Profile of Indirect Inelastic X-Ray Scattering in Correlated Materials: Observing Two-Magnon Scattering and Relation to the Dynamical Structure Factor}}},\ }\href@noop {} {\bibfield  {journal} {\bibinfo  {journal} {New J. Phys.}\ }\textbf {\bibinfo {volume} {14}},\ \bibinfo {pages} {113038} (\bibinfo {year} {2012})}\BibitemShut {NoStop}%
\bibitem [{\citenamefont {Jia}\ \emph {et~al.}(2014)\citenamefont {Jia}, \citenamefont {Nowadnick}, \citenamefont {Wohlfeld}, \citenamefont {Kung}, \citenamefont {Chen}, \citenamefont {Johnston}, \citenamefont {Tohyama}, \citenamefont {Moritz},\ and\ \citenamefont {Devereaux}}]{jia2014persistent}%
  \BibitemOpen
  \bibfield  {author} {\bibinfo {author} {\bibfnamefont {C.}~\bibnamefont {Jia}}, \bibinfo {author} {\bibfnamefont {E.}~\bibnamefont {Nowadnick}}, \bibinfo {author} {\bibfnamefont {K.}~\bibnamefont {Wohlfeld}}, \bibinfo {author} {\bibfnamefont {Y.}~\bibnamefont {Kung}}, \bibinfo {author} {\bibfnamefont {C.-C.}\ \bibnamefont {Chen}}, \bibinfo {author} {\bibfnamefont {S.}~\bibnamefont {Johnston}}, \bibinfo {author} {\bibfnamefont {T.}~\bibnamefont {Tohyama}}, \bibinfo {author} {\bibfnamefont {B.}~\bibnamefont {Moritz}},\ and\ \bibinfo {author} {\bibfnamefont {T.}~\bibnamefont {Devereaux}},\ }\bibfield  {title} {\bibinfo {title} {{\textit{Persistent Spin Excitations in Doped Antiferromagnets Revealed by Resonant Inelastic Light Scattering}}},\ }\href@noop {} {\bibfield  {journal} {\bibinfo  {journal} {Nat. Commun.}\ }\textbf {\bibinfo {volume} {5}},\ \bibinfo {pages} {3314} (\bibinfo {year} {2014})}\BibitemShut {NoStop}%
\bibitem [{bio()}]{biorthogonal}%
  \BibitemOpen
  \href@noop {} {}\bibinfo {note} {For general systems, constructing a dual Krylov subspace $\mathcal{L}_m$ alongside the primary Krylov subspace $\mathcal{K}_m$ is necessary, introducing additional MVMs and requiring the use of BiCG algorithms. However, when the Hamiltonian $\Ham$ is a real symmetric, as is typically the case, the two subspaces can be derived from one another with negligible computational overhead. As a result, within the scope of this paper, we do not explicitly distinguish between BiCG and CG/MINRES frameworks, since the algorithm choice naturally follows from the real valuedness and symmetry properties of the system. Further details are provided in the Supplementary Materials.}\BibitemShut {Stop}%
\bibitem [{\citenamefont {K{\"u}hner}\ and\ \citenamefont {White}(1999)}]{kuhner1999dynamical}%
  \BibitemOpen
  \bibfield  {author} {\bibinfo {author} {\bibfnamefont {T.~D.}\ \bibnamefont {K{\"u}hner}}\ and\ \bibinfo {author} {\bibfnamefont {S.~R.}\ \bibnamefont {White}},\ }\bibfield  {title} {\bibinfo {title} {{\textit{Dynamical Correlation Functions Using the Density Matrix Renormalization Group}}},\ }\href@noop {} {\bibfield  {journal} {\bibinfo  {journal} {Phys. Rev. B}\ }\textbf {\bibinfo {volume} {60}},\ \bibinfo {pages} {335} (\bibinfo {year} {1999})}\BibitemShut {NoStop}%
\bibitem [{\citenamefont {Nocera}\ and\ \citenamefont {Alvarez}(2016)}]{nocera2016spectral}%
  \BibitemOpen
  \bibfield  {author} {\bibinfo {author} {\bibfnamefont {A.}~\bibnamefont {Nocera}}\ and\ \bibinfo {author} {\bibfnamefont {G.}~\bibnamefont {Alvarez}},\ }\bibfield  {title} {\bibinfo {title} {{\textit{Spectral Functions with the Density Matrix Renormalization Group: Krylov-Space Approach for Correction Vectors}}},\ }\href@noop {} {\bibfield  {journal} {\bibinfo  {journal} {Phys. Rev. E}\ }\textbf {\bibinfo {volume} {94}},\ \bibinfo {pages} {053308} (\bibinfo {year} {2016})}\BibitemShut {NoStop}%
\bibitem [{\citenamefont {Kumar}\ \emph {et~al.}(2018)\citenamefont {Kumar}, \citenamefont {Nocera}, \citenamefont {Dagotto},\ and\ \citenamefont {Johnston}}]{kumar2018multi}%
  \BibitemOpen
  \bibfield  {author} {\bibinfo {author} {\bibfnamefont {U.}~\bibnamefont {Kumar}}, \bibinfo {author} {\bibfnamefont {A.}~\bibnamefont {Nocera}}, \bibinfo {author} {\bibfnamefont {E.}~\bibnamefont {Dagotto}},\ and\ \bibinfo {author} {\bibfnamefont {S.}~\bibnamefont {Johnston}},\ }\bibfield  {title} {\bibinfo {title} {\textit{Multi-Spinon and Antiholon Excitations Probed by Resonant Inelastic X-Ray Scattering on Doped One-Dimensional Antiferromagnets}},\ }\href@noop {} {\bibfield  {journal} {\bibinfo  {journal} {New J. Phys.}\ }\textbf {\bibinfo {volume} {20}},\ \bibinfo {pages} {073019} (\bibinfo {year} {2018})}\BibitemShut {NoStop}%
\bibitem [{\citenamefont {Nocera}\ \emph {et~al.}(2018)\citenamefont {Nocera}, \citenamefont {Kumar}, \citenamefont {Kaushal}, \citenamefont {Alvarez}, \citenamefont {Dagotto},\ and\ \citenamefont {Johnston}}]{nocera2018computing}%
  \BibitemOpen
  \bibfield  {author} {\bibinfo {author} {\bibfnamefont {A.}~\bibnamefont {Nocera}}, \bibinfo {author} {\bibfnamefont {U.}~\bibnamefont {Kumar}}, \bibinfo {author} {\bibfnamefont {N.}~\bibnamefont {Kaushal}}, \bibinfo {author} {\bibfnamefont {G.}~\bibnamefont {Alvarez}}, \bibinfo {author} {\bibfnamefont {E.}~\bibnamefont {Dagotto}},\ and\ \bibinfo {author} {\bibfnamefont {S.}~\bibnamefont {Johnston}},\ }\bibfield  {title} {\bibinfo {title} {\textit{Computing Resonant Inelastic X-Ray Scattering Spectra using the Density Matrix Renormalization Group Method}},\ }\href@noop {} {\bibfield  {journal} {\bibinfo  {journal} {Sci. Rep.}\ }\textbf {\bibinfo {volume} {8}},\ \bibinfo {pages} {11080} (\bibinfo {year} {2018})}\BibitemShut {NoStop}%
\bibitem [{\citenamefont {van~der Vorst}\ and\ \citenamefont {Melissen}(1990)}]{van1990petrov}%
  \BibitemOpen
  \bibfield  {author} {\bibinfo {author} {\bibfnamefont {H.~A.}\ \bibnamefont {van~der Vorst}}\ and\ \bibinfo {author} {\bibfnamefont {J.~B.}\ \bibnamefont {Melissen}},\ }\bibfield  {title} {\bibinfo {title} {{\textit{A Petrov-Galerkin Type Method for Solving Ak = B, Where A Is Symmetric Complex}}},\ }\href@noop {} {\bibfield  {journal} {\bibinfo  {journal} {IEEE Transactions on Magnetics}\ }\textbf {\bibinfo {volume} {26}},\ \bibinfo {pages} {706} (\bibinfo {year} {1990})}\BibitemShut {NoStop}%
\bibitem [{\citenamefont {Frommer}(2003)}]{frommer2003BiCGStab}%
  \BibitemOpen
  \bibfield  {author} {\bibinfo {author} {\bibfnamefont {A.}~\bibnamefont {Frommer}},\ }\bibfield  {title} {\bibinfo {title} {{\textit{BiCGStab($l$) for Families of Shifted Linear Systems}}},\ }\href@noop {} {\bibfield  {journal} {\bibinfo  {journal} {Computing}\ }\textbf {\bibinfo {volume} {70}},\ \bibinfo {pages} {87–109} (\bibinfo {year} {2003})}\BibitemShut {NoStop}%
\bibitem [{\citenamefont {Meng}\ and\ \citenamefont {Li}(2015)}]{meng2015recycling}%
  \BibitemOpen
  \bibfield  {author} {\bibinfo {author} {\bibfnamefont {J.}~\bibnamefont {Meng}}\ and\ \bibinfo {author} {\bibfnamefont {H.}~\bibnamefont {Li}},\ }\bibfield  {title} {\bibinfo {title} {{\textit{Recycling Bicg for Families of Shifted Linear Systems}}},\ }in\ \href@noop {} {\emph {\bibinfo {booktitle} {2015 11th International Conference on Computational Intelligence and Security (CIS)}}}\ (\bibinfo {organization} {IEEE},\ \bibinfo {year} {2015})\ p.~\bibinfo {pages} {86}\BibitemShut {NoStop}%
\bibitem [{\citenamefont {Lehoucq}\ \emph {et~al.}(1998)\citenamefont {Lehoucq}, \citenamefont {Sorensen},\ and\ \citenamefont {Yang}}]{lehoucq1998arpack}%
  \BibitemOpen
  \bibfield  {author} {\bibinfo {author} {\bibfnamefont {R.~B.}\ \bibnamefont {Lehoucq}}, \bibinfo {author} {\bibfnamefont {D.~C.}\ \bibnamefont {Sorensen}},\ and\ \bibinfo {author} {\bibfnamefont {C.}~\bibnamefont {Yang}},\ }\href@noop {} {\emph {\bibinfo {title} {{\textit{{Arpack Users' Guide: Solution of Large-Scale Eigenvalue Problems with Implicitly Restarted Arnoldi Methods}}}}}}\ (\bibinfo  {publisher} {SIAM},\ \bibinfo {year} {1998})\BibitemShut {NoStop}%
\bibitem [{\citenamefont {Jia}\ \emph {et~al.}(2018)\citenamefont {Jia}, \citenamefont {Wang}, \citenamefont {Mendl}, \citenamefont {Moritz},\ and\ \citenamefont {Devereaux}}]{jia2017paradeisos}%
  \BibitemOpen
  \bibfield  {author} {\bibinfo {author} {\bibfnamefont {C.}~\bibnamefont {Jia}}, \bibinfo {author} {\bibfnamefont {Y.}~\bibnamefont {Wang}}, \bibinfo {author} {\bibfnamefont {C.}~\bibnamefont {Mendl}}, \bibinfo {author} {\bibfnamefont {B.}~\bibnamefont {Moritz}},\ and\ \bibinfo {author} {\bibfnamefont {T.}~\bibnamefont {Devereaux}},\ }\bibfield  {title} {\bibinfo {title} {{\textit{{Paradeisos: a Perfect Hashing Algorithm for Many-Body Eigenvalue Problems}}}},\ }\href@noop {} {\bibfield  {journal} {\bibinfo  {journal} {Comput. Phys. Commun.}\ }\textbf {\bibinfo {volume} {224}},\ \bibinfo {pages} {81} (\bibinfo {year} {2018})}\BibitemShut {NoStop}%
\bibitem [{\citenamefont {Betts}\ \emph {et~al.}(1999)\citenamefont {Betts}, \citenamefont {Lin},\ and\ \citenamefont {Flynn}}]{betts1999improved}%
  \BibitemOpen
  \bibfield  {author} {\bibinfo {author} {\bibfnamefont {D.}~\bibnamefont {Betts}}, \bibinfo {author} {\bibfnamefont {H.}~\bibnamefont {Lin}},\ and\ \bibinfo {author} {\bibfnamefont {J.}~\bibnamefont {Flynn}},\ }\bibfield  {title} {\bibinfo {title} {{\textit{Improved Finite-Lattice Estimates of the Properties of Two Quantum Spin Models on the Infinite Square Lattice}}},\ }\href@noop {} {\bibfield  {journal} {\bibinfo  {journal} {Can. J. Phys.}\ }\textbf {\bibinfo {volume} {77}},\ \bibinfo {pages} {353} (\bibinfo {year} {1999})}\BibitemShut {NoStop}%
\bibitem [{\citenamefont {Wang}\ \emph {et~al.}(2020)\citenamefont {Wang}, \citenamefont {He}, \citenamefont {Wohlfeld}, \citenamefont {Hashimoto}, \citenamefont {Huang}, \citenamefont {Lu}, \citenamefont {Mo}, \citenamefont {Komiya}, \citenamefont {Jia}, \citenamefont {Moritz} \emph {et~al.}}]{wang2020emergence}%
  \BibitemOpen
  \bibfield  {author} {\bibinfo {author} {\bibfnamefont {Y.}~\bibnamefont {Wang}}, \bibinfo {author} {\bibfnamefont {Y.}~\bibnamefont {He}}, \bibinfo {author} {\bibfnamefont {K.}~\bibnamefont {Wohlfeld}}, \bibinfo {author} {\bibfnamefont {M.}~\bibnamefont {Hashimoto}}, \bibinfo {author} {\bibfnamefont {E.~W.}\ \bibnamefont {Huang}}, \bibinfo {author} {\bibfnamefont {D.}~\bibnamefont {Lu}}, \bibinfo {author} {\bibfnamefont {S.-K.}\ \bibnamefont {Mo}}, \bibinfo {author} {\bibfnamefont {S.}~\bibnamefont {Komiya}}, \bibinfo {author} {\bibfnamefont {C.}~\bibnamefont {Jia}}, \bibinfo {author} {\bibfnamefont {B.}~\bibnamefont {Moritz}}, \emph {et~al.},\ }\bibfield  {title} {\bibinfo {title} {{\textit{Emergence of Quasiparticles in a Doped Mott Insulator}}},\ }\href@noop {} {\bibfield  {journal} {\bibinfo  {journal} {Commun. Phys.}\ }\textbf {\bibinfo {volume} {3}},\ \bibinfo {pages} {210} (\bibinfo {year} {2020})}\BibitemShut {NoStop}%
\bibitem [{\citenamefont {Van~den Brink}\ and\ \citenamefont {Van~Veenendaal}(2005)}]{van2005correlation}%
  \BibitemOpen
  \bibfield  {author} {\bibinfo {author} {\bibfnamefont {J.}~\bibnamefont {Van~den Brink}}\ and\ \bibinfo {author} {\bibfnamefont {M.}~\bibnamefont {Van~Veenendaal}},\ }\bibfield  {title} {\bibinfo {title} {{\textit{Correlation Functions Measured by Indirect Resonant Inelastic X-Ray Scattering}}},\ }\href@noop {} {\bibfield  {journal} {\bibinfo  {journal} {Europhys. Lett.}\ }\textbf {\bibinfo {volume} {73}},\ \bibinfo {pages} {121} (\bibinfo {year} {2005})}\BibitemShut {NoStop}%
\bibitem [{BiC()}]{BiCGStab}%
  \BibitemOpen
  \href@noop {} {}\bibinfo {note} {The BiCGStab algorithm enhances the standard BiCG approach by introducing a gradient stabilization mechanism, accelerating convergence and further reducing the actual speedup from the ideal $2n$. To ensure a fair benchmark against the best-performing algorithms currently available for resonant spectral simulations, we compare MSBiCG with BiCGStab throughout this paper.}\BibitemShut {Stop}%
\bibitem [{\citenamefont {Sharma}\ \emph {et~al.}()\citenamefont {Sharma} \emph {et~al.}}]{data}%
  \BibitemOpen
  \bibfield  {author} {\bibinfo {author} {\bibfnamefont {P.}~\bibnamefont {Sharma}} \emph {et~al.}\ }\href {https://doi.org/10.6084/m9.figshare.29972677.v1} {10.6084/m9.figshare.29972677.v1}\BibitemShut {NoStop}%
\end{thebibliography}%

\end{document}